\documentclass[onecolumn,superscriptaddress,amsmath,aps,pra]{revtex4}

\usepackage{graphicx}
\usepackage{wrapfig}
\usepackage{appendix}
\usepackage{subfigure}
\usepackage{textcomp}

\begin{document}

\title{Layer by layer generation of cluster states}
\author{Katherine L. Brown}
 \thanks{Current address: Hearne Institute for Theoretical Physics and Department of Physics and Astronomy, Louisiana State University, Baton Rouge, LA, 70803, USA; katherineb@lsu.edu}
   \affiliation{School of Physics and Astronomy, University of Leeds, Leeds, LS2 9JT, UK}
  \author{Clare Horsman}
   \thanks{Current address: Keio University Shonan Fujisawa Campus, Fujisawa, Kanagawa 252-0882, Japan}
   \affiliation{Department of Mathematics, University of Bristol, University Walk, Bristol BS8 1TW, UK}  
   \affiliation{H. H. Wills Physics Laboratory, University of Bristol, Tyndall Avenue, Bristol, BS8 1TL, UK}
\author{Viv Kendon}
   \affiliation{School of Physics and Astronomy, University of Leeds, Leeds, LS2 9JT, UK}
\author{William J. Munro}
 \affiliation{NTT Basic Research Laboratories, NTT Corporation, 3-1 Morinosato-Wakamiya, Atsugi-shi, Kanagawa-ken 243-0198, Japan}   
   \affiliation{National Institute of Informatics, 2-1-2 Hitotsubashi, Chiyoda-ku, Tokyo 101-8430, Japan}
   
\date{\today}
   
\begin{abstract}    
Cluster states can be used to perform measurement-based quantum computation. The cluster state is a useful resource, because once it has been generated only local operations and measurements are needed to perform universal quantum computation. In this paper, we explore techniques for quickly and deterministically building a cluster state. In particular we consider generating cluster states on a qubus quantum computer, a computational architecture which uses a continuous variable ancilla to generate interactions between qubits. We explore several techniques for building the cluster, with the number of operations required depending on whether we allow the ability to destroy previously created controlled-phase links between qubits. In the case where we can not destroy these links, we show how to create an $n \times m$ cluster using just $3nm -2n -3m/2 + 3$ operations. This gives more than a factor of 2 saving over a na\"ive method. Further savings can be obtained if we include the ability to destroy links, in which case we only need $\frac{1}{3}(8nm-4n-4m-8)$ operations. Unfortunately the latter scheme is more complicated so choosing the correct order to interact the qubits is considerably more difficult. A half way scheme, that keeps a modular generation but saves additional operations over never destroying links requires only $3nm-2n-2m+4$ operations. The first scheme and the last scheme are the most practical for building a cluster state because they split up the generation into the repetition of simple sections. 
\end{abstract}

\maketitle

\section{Introduction} \label{sec:introduction}
Quantum computation is a field of research which uses quantum mechanics rather than classical mechanics to build a computer \cite{Nielsen2000}. When built, a large scale quantum computer will give significant efficiency advantages over a classical computer. The most famous example of an improvement is Shor's factoring algorithm, which gives an exponential reduction in the time required compared to the best known classical algorithm \cite{Shor1997}. However, building a quantum computer is not easy, as the level of control required makes it very challenging to build anything but the smallest systems. This is problematic as the break even point (the point where a quantum computer is more efficient than a classical computer) for Shor's algorithm is in the region of thousands of logical qubits \cite{VanMeter2006}. Even for promising applications, such as quantum simulation, a universal quantum simulator which can simulate an arbitrary system, will require hundreds of qubits to get a significant improvement over current classical methods \cite{Brown2010}.

One proposal to get around some of the control difficulties is that of using cluster states for quantum computing \cite{Raussendorf2001,Raussendorf2003}. A cluster state can be represented as a lattice-graph, with qubits as the nodes and controlled phase (C-Phase) gates as the edges. Once this highly entangled resource has been created, only measurements and simple local unitary corrections need to be applied, in order to obtain universal quantum computing. This system provides significant advantages, as all the entanglement takes the form of C-Phase gates, so removes the need for many different entangling gates or a large number of local unitaries. Another advantage is that all the entanglement is generated off-line so any corrections can be performed before the quantum algorithm begins. 

Many schemes for generating cluster states either assume a C-Phase gate can be realised, or concentrate on probabilistic gates, and the best methods of generating the state given these operations  \cite{Nielsen2004,Browne2005,Gross2006,Gilbert2006,Kieling2007,Kok2007,Louis2007}. While the original discussion \cite{Raussendorf2001} of cluster states suggested a one-shot entangling operation this hasn't been realised experimentally. Other schemes which have used a fixed number of global operations are equally difficult to implement experimentally \cite{Borhani2005,Wei2011}. Although probabilistic schemes can, in theory, grow large clusters, there are significant problems with them in practise. Probabilistic gates with low success probabilities ($\sim 50\%$) dramatically increase the number of operations required, compared to deterministic operations. Given that decoherence remains a major problem in all qubit systems, this increase in time could be detrimental when it comes to creating a large cluster. Similarly, although these schemes are heralded, they would require complex routing structures dependent upon which operations were successful. While schemes have been suggested to avoid some of these problems \cite{Campbell2007,Kieling2007,Matsuzaki2010} a deterministic scheme would still be useful. 

In this paper we describe a scheme to use a form of ancilla-based computation, called the qubus \cite{Loock2008,Spiller2006}, to generate cluster states. This qubus scheme uses a `quantum bus' in the form of a continuous variable field mode to generate interactions between qubits \cite{Loock2008}. It has been previously shown that such a scheme can be used to generate a deterministic C-Phase gate \cite{Spiller2006}. Another significant advantage of using this system is that we never need the qubits to interact directly with each other. Entangling qubits via direct interaction often means they need to be placed too close together to give the individual addressability necessary to implement algorithms on the cluster state \cite{Jane2003}. Furthermore, it is important that we create the cluster in a dynamic fashion. This means that we can use the earliest parts of the cluster while still generating the later parts \cite{Horsman2010}. This avoids the early parts of the cluster decohering before we get the chance to perform operations on them. The new schemes we present in this paper are dynamic. Finally while we concentrate here on building cluster states using the qubus quantum computer, with minimal adaptation our results will also be applicable to other ancilla-based systems. 

This paper is structured as follows: we begin by introducing the qubus architecture in section \ref{sec:qubus}, where we also give a brief description of how to save operations over a na\"ive method of implementation.  In section \ref{Sec:Limitations} we discuss the largest sections of the cluster we can create, interacting each qubit with the bus only twice (once to entangle the qubit with the bus, and once to disentangle the qubit from the bus), assuming that we don't want to create then destroy interactions between our qubits. This allows us to place a bound on the size of the layers we will use for building our cluster. In section \ref{Simple} we show how to combine these layers in a simple fashion, and thus calculate how many operations are required to create a large cluster, in the simplest case. By keeping some qubits entangled with the bus between layers it is possible to save operations, and we discuss the techniques necessary to achieve this in section \ref{Comb}. We summarise this work in section \ref{Tot}, and show that these schemes still use more operations than the lower bounds we obtained in a previous paper \cite{Horsman2010}. In section \ref{Neg} we look at what happens if we relax our previous conditions; in particular we allow links between the qubits to be created and destroyed when transitioning between layers. This is significant as it allows us achieve the lower bound that we obtained in \cite{Horsman2010}. However, in section \ref{Rev} we show that our previous work on the limitations of the bus becomes invalid in this case, and we can use larger layers. This allows us to generate clusters using even fewer operations than our previous lower bound. However, as we increase the size of the layers, the complexity of our scheme increases, and we also lose the dynamic nature of our generation scheme. Finally we summarise and conclude in section \ref{Conclusions}. 

\section{The Qubus}\label{sec:qubus}
The qubus quantum computer is a hybrid architecture, consisting of a continuous variable field mode (referred to as the bus) generating entanglement between the qubits of the main data processing system \cite{Nemoto2004, Munro2005, Spiller2006}. The advantage of such a scheme is that it removes the need for qubits to interact directly with each other. This enables distant qubits to be entangled without using swap operations to bring them next to each other. In a system of nearest neighbour interactions, the removal of the requirement for direct interaction between the qubits takes away the need to change the calibration settings every time a new qubit is added. Our qubits can therefore be placed at a reasonable distance from each other, removing any difficulties which may occur with individual addressibility \cite{Jane2003}. Equivalent schemes which use single photons as a means of generating entanglement between qubits have also been proposed; however these involve the currently experimentally infeasible process of creating and detecting single photons reliably \cite{Cirac1997, Mancini2004, Lim2005, Duan2005, Barrett2005,Dixon2009}. A qubus scheme removes this difficulty by eliminating the need for photon detection completely. 

We consider generating interactions by performing displacement operations, conditional on the state of the qubits, on one of the field quadratures of the bus. The displacement operator takes the form $D(\beta \sigma_{zk}) = \exp(\beta \sigma_{zk} a^\dagger - \beta^* \sigma_{zk} a)$, where $a^{\dagger}$, $a$ are the creation and annihilation operators, $\sigma_{zk}$ is the Pauli Z operator acting on the $k^{\text{th}}$ qubit, and  $\beta = \chi t e^{i(\phi - \frac{\pi}{2})}$, where $t$ is time, and $\chi$ is the strength of the non-linearity being used \cite{Spiller2006}. In the present case $\phi=0$ (corresponding to the position quadrature) or $\phi = \frac{\pi}{2}$ (corresponding to the momentum quadrature). The interactions between nearest neighbour sites in a cluster state take the form of the maximally entangling C-Phase interaction. It is possible to generate the C-Phase interaction on the qubus deterministically using conditional displacement gates. If we have an interaction Hamiltonian of the form
\begin{equation} \label{eq:interaction1}
H_{\text{int}} = \hbar\chi\sigma_{z}(a^{\dagger}e^{i\theta} + ae^{-i\theta})
\end{equation}
interacting for a time $t$, then this performs a displacement operator of the form $D(\sigma_{z}\beta)$ on the field \cite{Spiller2006}. If the controlled displacement is not available directly, then we can create the controlled displacement through controlled rotations and unconditional displacements \cite{Loock2008}. To make a single C-Phase gate, we will consider a scheme consisting of two qubits and a bus. The first qubit is coupled to the momentum quadrature of the bus [i.e. $\theta = \pi/2$ in equation (\ref{eq:interaction1})], and the second qubit is coupled to the position quadrature of the bus [i.e. $\theta=0$ in equation (\ref{eq:interaction1})]. These couplings will result in displacements in orthogonal directions, allowing us to combine them using the formula  
\begin{equation} \label{combinedisplacement}
D(a)D(b) = \exp \left[ \frac{ab^{*} - a^{*}b}{2} \right] D(a + b)
\end{equation}
where $a$ and $b$ are either complex numbers or commuting operators which act on the qubits. 

\begin{figure}
\includegraphics[width=12cm]{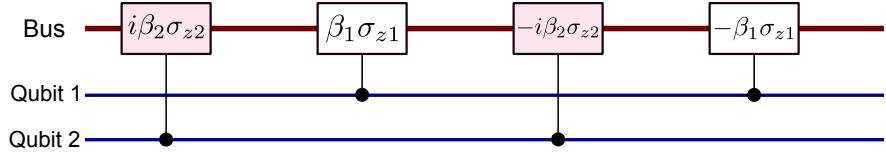}
\caption{A circuit diagram showing how to perform a maximally entangling gate on the qubus. The shaded boxes are displacement operators on the position quadrature of the bus, while the white boxes are displacement operators on the momentum quadrature of the bus.}  \label{cphase}
\end{figure}

It is possible to create a C-Phase gate using four displacement operators, two of which entangle the qubits with the bus and two of which disentangle the qubits from the bus. Such a scheme is illustrated in figure \ref{cphase}. 
The displacement operators required to perform a two qubit gate, $U_{\text{cp}}$, that is local unitaries away from the controlled phase gate are
\begin{equation}
U_{\text{cp}} = D(i\beta_{2}\sigma_{z2}) D(\beta_{1}\sigma_{z1})D(-i\beta_{2}\sigma_{z2})D(-\beta_{1}\sigma_{z1})
\end{equation}
where $\beta_{1}$ and $\beta_{2}$ are real phase factors. Operating this on two qubits in an equal superposition of $|0\rangle $ and $|1\rangle$ gives
\begin{equation}
|\psi\rangle = \frac{1}{2} U_{\text{cp}} \left(|0\rangle + |1\rangle\right) \left(|0\rangle + |1\rangle \right) = \frac{1}{2} \exp(2i\beta_{1}\beta_{2}\sigma_{z1}\sigma_{z2}) \left(|0\rangle + |1\rangle\right) \left(|0\rangle + |1\rangle \right)
\end{equation}
If we set $2\beta_{1}\beta{2}=\pi/4$ then we get the state
\begin{equation}
|\psi\rangle =  \frac{1}{2} \left(e^{i\frac{\pi}{4}} |00\rangle + e^{-i\frac{\pi}{4}}|01\rangle + e^{-i\frac{\pi}{4}}|10\rangle + e^{i\frac{\pi}{4}}|11\rangle \right)
\end{equation}
Up to local unitaries, of the form $\{\exp(-i\pi/8)|0\rangle \langle 0 | + \exp(3i\pi/8)|1\rangle \langle 1|\}$, this is now the outcome of the conventional CPhase gate:
\begin{equation}
|\psi\rangle =  \frac{1}{2} \left( |00\rangle + |01\rangle + |10\rangle - |11\rangle \right)
\end{equation}
At the end of this sequence of displacement operators, the bus will be completely disentangled from the qubits. This can be seen from that fact that no net displacement operator will have been performed on the bus. Provided the bus is disentangled at the end of a sequence of displacement operators, it will be possible to perform another sequence of operators without generating any unwanted entanglement. If we have two sequences of displacement operators, all of which commute, and in which some terms are repeated multiple times, it is possible to combine these two sequences, reducing the total number of displacement operators required. An example is replacing a sequence of operators given by 
\begin{equation}
U_{F}=D(\beta_{1}\sigma_{z1})D(-i\beta_{2}\sigma_{z2})D(-\beta_{1}\sigma_{z1})D(i\beta_{2}\sigma_{z2})D(\beta_{1}\sigma_{z1})D(-i\beta_{3}\sigma_{z3})D(-\beta_{1}\sigma_{z1})D(i\beta_{3}\sigma_{z3})
\end{equation}
with the set
\begin{equation}
U_{R} = D(\beta_{1}\sigma_{z1})D(-i\beta_{2}\sigma_{z2})D(-i\beta_{3}\sigma_{z3})D(-\beta_{1}\sigma_{z1})D(i\beta_{2}\sigma_{z2})D(i\beta_{3}\sigma_{z3})
\end{equation}
Both of these sequences generate an interaction between the qubits of the form 
\begin{equation}
U_{F} = U_{R} = \exp(2i\beta_{1}\beta_{2}\sigma_{z1}\sigma_{z2}+2i\beta_{1}\beta_{3}\sigma_{z1}\sigma_{z3})
\end{equation}
The sequence $U_{R}$ requires only 6 operations, while the sequence $U_{F}$ requires 8 operations. This is the simplest reduction possible, and when we consider entanglement between more qubits we can get further savings.  

In this paper, we will use similar techniques to reduce the number of operations required for generating cluster states. We will show that it is possible to get a considerable reduction in the number of interactions with the bus compared to a na\"ive method of implementing each C-Phase gate individually. The efficiency of generating cluster states on a qubus quantum computer was previously considered by Louis et al.~\cite{Louis2007}. Of particular interest to this current work is the measurement-free scheme, which they only briefly touched on. In their paper, Louis et al.~showed that it was possible to build a cluster state by entangling the qubits in chains, first in the horizontal and then in the vertical direction. A chain of $n$ qubits could be entangled using just $2n$ operations, therefore an $n \times m$ cluster could be entangled using just $4nm$ operations. This showed significant improvements compared to a na\"ive method where $8nm -4m -4n$ operations are required to generate each C-Phase gate in the cluster individually. 

In previous work we showed how to get a saving in the number of operations required when using the qubus for simulating the BCS Hamiltonian \cite{Brown2011}. We found it was possible to get up to O$(n)$ savings in generating entanglement between a fully connected line of qubits. The nearest neighbour case of the BCS Hamiltonian is equivalent to a spin-chain or a 1D cluster state. We now build upon this previous work and look at how to get savings in a 2D case rather than just for a 1D chain. While Louis et al.~\cite{Louis2007} considered generating a cluster from 1D chains, we will use a base unit that is a 2D section of our cluster, and thereby obtaining further savings. We have explored this previously in \cite{Horsman2010} where we focused on error accumulation when using the qubus to generate cluster states. In our current paper we show techniques for generating our cluster using a minimum number of operations, and explore how allowing the destruction of operations effects the total number of operations needed to build a cluster. 

\section{The limitations of the bus} \label{Sec:Limitations}
We now consider how we can generate an $n \times m$ cluster in a layered fashion using our qubus quantum computer. We will consider using layers which have a length $n$, and a fixed width of 2, 3 or 4 qubits. By stacking these layers together to form a total width of $m$, it is possible to build our entire cluster. This gives us a convenient technique for counting how many operations we need to build a cluster. We will consider using layers which can be built by interacting each qubit with the bus only twice, once to entangle the qubit with the bus, and once to remove it. In our diagrams, pink qubits will be connected to the momentum quadrature or the bus, and blue qubits to the position quadrature. Owing to the nature of the qubus we never generate interactions between qubits of the same colour. We will also begin by placing a limitation that we don't want to create, then destroy a link. This avoids additional errors that will occur if two links aren't performed exactly. We will later relax this restriction, to see what further improvements are possible. In our previous work \cite{Horsman2010} we looked at placing a lower bound on the total number of operations required, given the inability to destroy links. By using a path across an $n \times m$ cluster, we suggested that this bound was $3nm - 2(n+m) -4$, and showed that it was possible to meet this lower bound using a spiral path containing only right-angle turns. 

\begin{figure}
\centering
\subfigure[]
{\includegraphics[width=5cm]{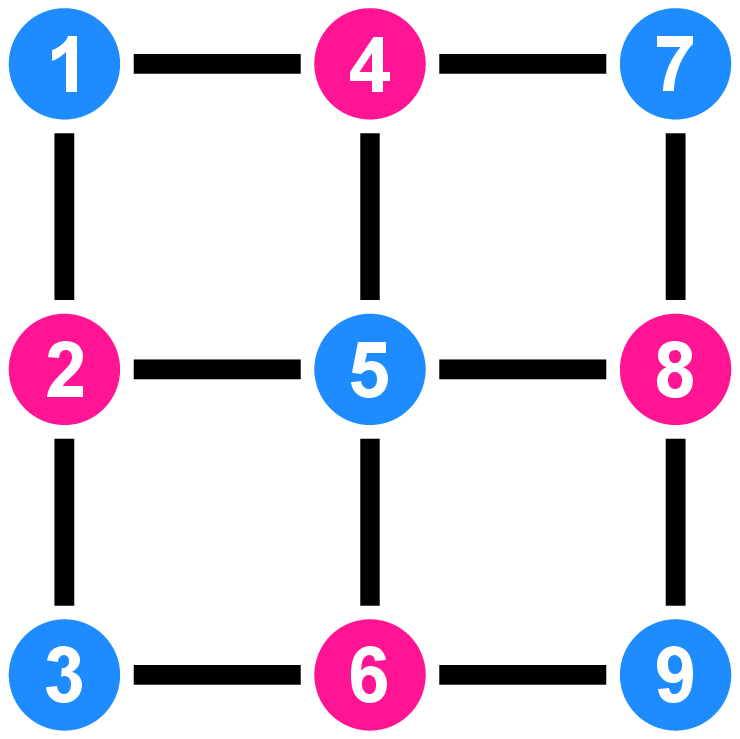}
\label{Fig:Gen:box}}
\subfigure[]
{\includegraphics[width=5cm]{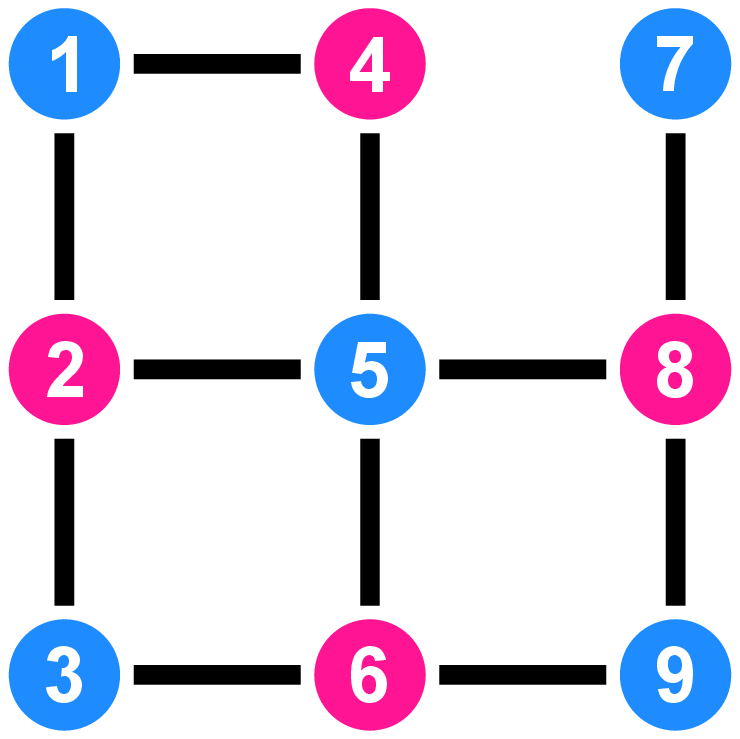}
\label{Fig:Gen:obox}}
\caption{Two shapes of cluster. (a) cannot be created using just one pair of interactions with the bus per qubit  however this is possible for (b).} 
\end{figure}

\begin{figure}
\centering
\subfigure[]
{\includegraphics[width=3.5cm]{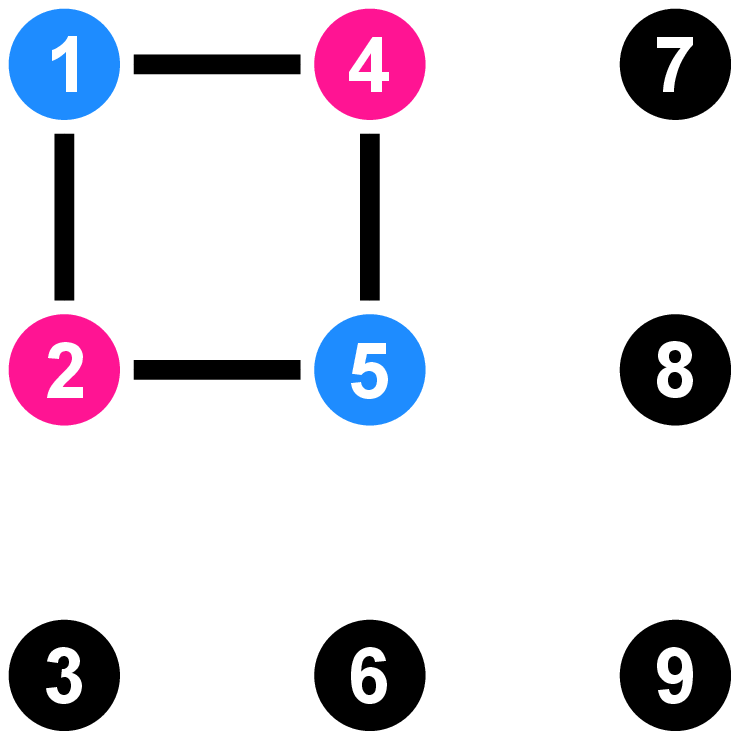}
\label{Fig:Gen:corner1}}
\subfigure[]
{\includegraphics[width=3.5cm]{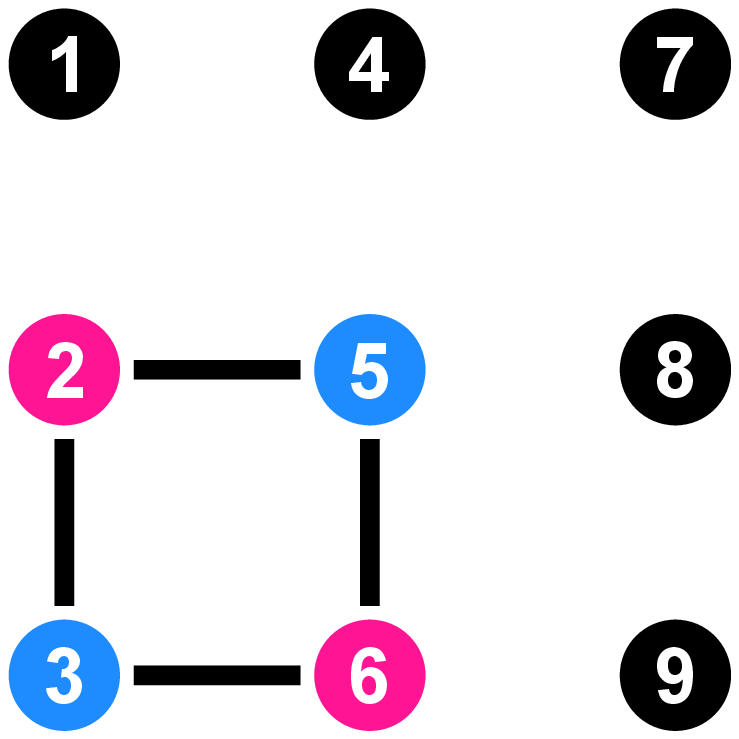}
\label{Fig:Gen:corner2}}
\subfigure[]
{\includegraphics[width=3.5cm]{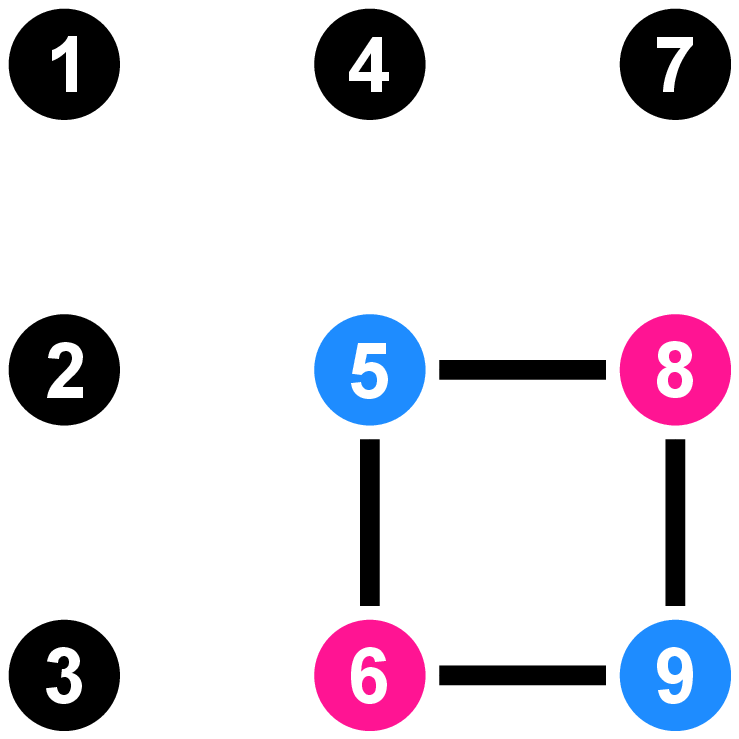}
\label{Fig:Gen:corner3}}
\subfigure[]
{\includegraphics[width=3.5cm]{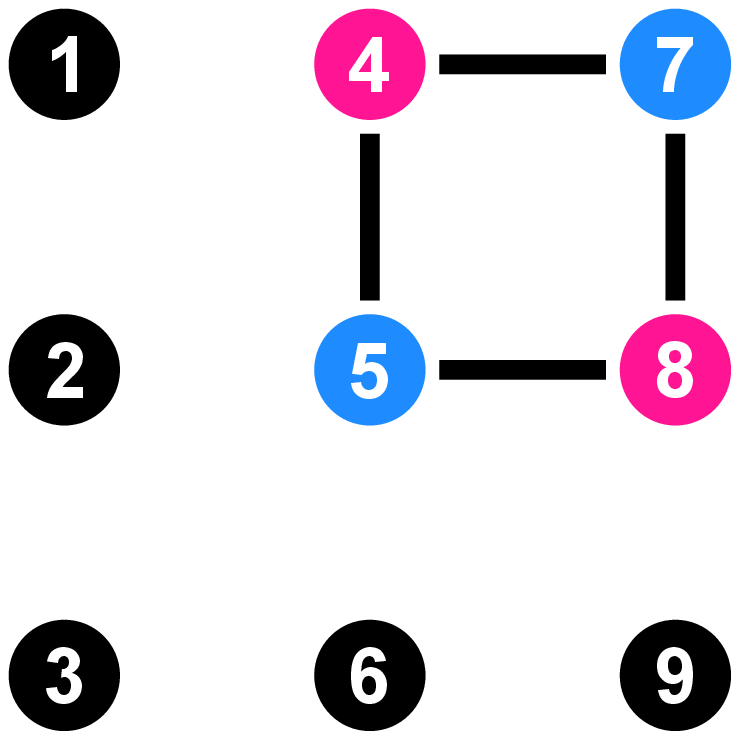}
\label{Fig:Gen:corner4}}
\caption{To make a $3\times3$ section of cluster we need to create the individual sections shown in (a),(b),(c) and (d). The black lines represent C-Phase gates between the qubits, we only generate operations between different coloured qubits.}\label{Fig:Gen:corners}
\end{figure}

\begin{figure}
\includegraphics[width=10cm]{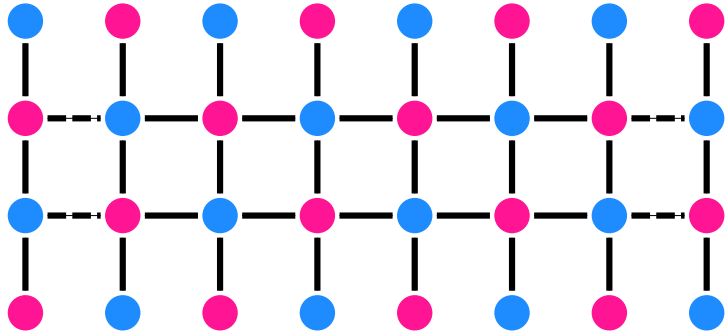}
\caption{The largest section of cluster it is possible to create without interacting qubits with the bus more than twice. The black lines represent C-Phase gates between the qubits, we only generate operations between different coloured qubits.}
\label{Layer4}
\end{figure}

We want to consider creating sealed sections of our cluster, to demonstrate what we can create given the restrictions we have placed on how we generate interactions. The simplest two dimensional example is a box of four qubits. In this case it is easy to create the necessary interactions using a set of operations given by 
\begin{equation}
U_{b} = D(\beta_{1}\sigma_{z1})D(-i\beta_{2}\sigma_{z2})D(-i\beta_{3}\sigma_{z3})D(-\beta_{1}\sigma_{z1})D(i\beta_{2}\sigma_{z2})D(i\beta_{3}\sigma_{z3})D(\beta_{4}\sigma_{z4})
\end{equation}
While we can create this box of 4 qubits, we can't scale this up and create a box of 9 qubits without interacting one qubit with the bus twice. This 9 qubit box is shown in figure \ref{Fig:Gen:box}. To create the box, we consider splitting it into four sections shown in figure \ref{Fig:Gen:corners}. Adjacent sections contain a common edge, and two common qubits. The common edge is not being created twice, rather it is repeated in two diagrams to represent the fact that the necessary qubits to create that edge are entangled with the bus for both diagrams. The operation sequence required for generating the sections of the cluster shown in figures \ref{Fig:Gen:corner1} and \ref{Fig:Gen:corner2} is
\begin{equation}
D(\beta_{1}\sigma_{z1})D(-i\beta_{2}\sigma_{z2})D(-i\beta_{4}\sigma_{z4})D(-\beta_{1}\sigma_{z1})D(-\beta_{5}\sigma_{z5})D(i\beta_{4}\sigma_{z4})D(-\beta_{3}\sigma_{z3})D(i\beta_{2}\sigma_{z2})D(i\beta_{6}\sigma_{z6}) \dots
\end{equation}
This illustrates the difficulty with creating our 9 qubit box without reactivating any qubits. Both diagram \ref{Fig:Gen:corner1} and \ref{Fig:Gen:corner4} require an edge which connects to qubit 4. While the edge between qubits 4 and 5 is common between the diagrams, so could be created in either, diagram \ref{Fig:Gen:corner1} contains an edge between qubits 1 and 4, while diagram \ref{Fig:Gen:corner4} contains an edge between qubits 4  and 7. This would require qubit 4 to be activated in both diagrams, or held on the bus between the two. The latter option is unfeasible as it would result in qubit 4 being entangled with qubit 3 and 9, thus destroying our cluster. This would be a significant disadvantage, as too-entangled systems are not suitable for use for measurement-based quantum computation \cite{Gross2009}. An alternative would be to generate the operations in figure \ref{Fig:Gen:corner4} immediately after the operations in figure \ref{Fig:Gen:corner1}. However, this would simply change the order in which we generate the sections of our box, and the same problem would occur when generating section \ref{Fig:Gen:corner2}. The largest section of cluster we can create in our $3 \times 3$ grid is shown in figure \ref{Fig:Gen:obox}. When we consider splitting our cluster into layers we find that the largest layer we can create is an open layer of width 4, which is illustrated in figure \ref{Layer4}. It is possible to build our cluster using this, or smaller layers, that can be represented by a portion of our largest possible layer. In the next section we will discuss techniques for combining layers of various sizes to find the best way to build our cluster.  

\section{A simple method for combining layers} \label{Simple}
\begin{figure}[tb]
\subfigure[]
{\includegraphics[width=7cm]{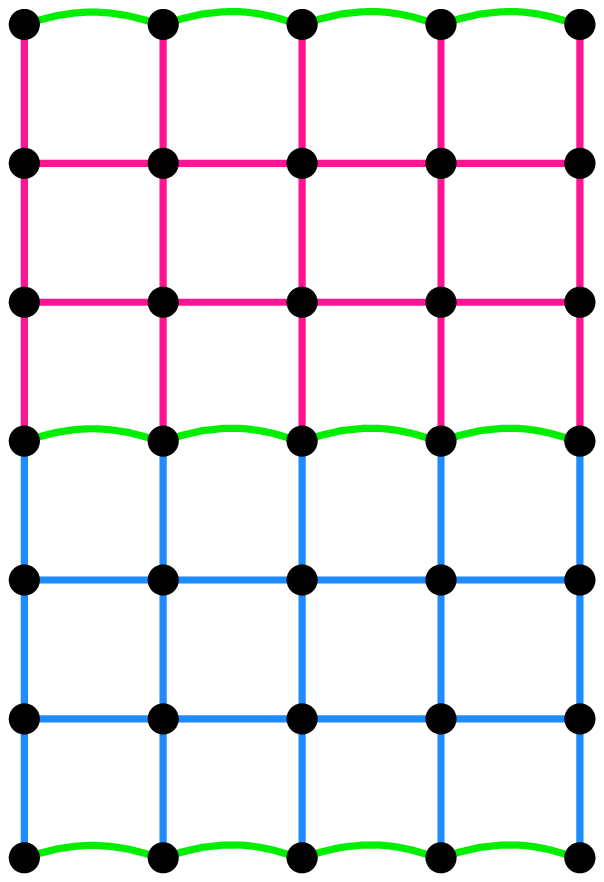}
\label{Fig:Gen:4box1}}
\subfigure[]
{\includegraphics[width=7cm]{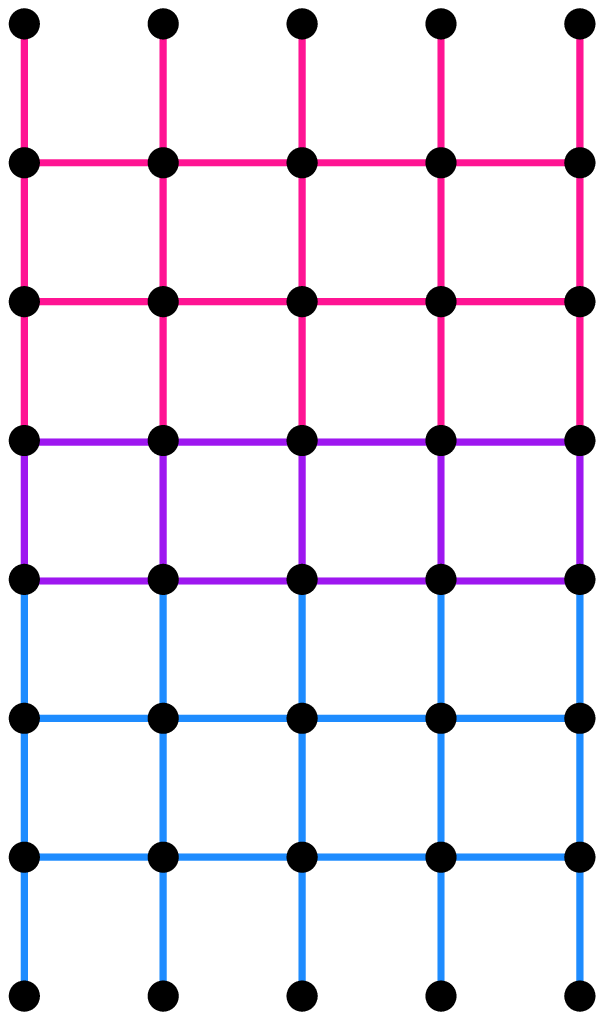}
\label{Fig:Gen:4box2}}
\caption{Two techniques to generate a cluster using a the layers of width 4. (a) involves stitching the qubits together using a path of C-Phase gates between single qubits. (b) involves alternating between a path of width 4 and a path of width 2.} 
\end{figure}

We now move on to consider what sized layers we want to use, and the best method to combine them. We start by considering using layers of width 4 stitched together at the ends, as illustrated in figure \ref{Fig:Gen:4box1}.  To work out the total number of operations required, we need to consider how many qubits interact with the bus more than once. Every qubit on the join of two sections has to interact with the bus three times: once as part of each layer, and once as part of the stitching. Each of these qubits therefore requires 4 operations with the bus, above the basic 2 required for all qubits. Since each layer has width 4, the number of qubits that are between two layers is given by $n(m-4)/3 $. An additional $4n$ operations are required to seal the top and bottom of the cluster. The total number of operations, above the basic $2nm$ required for generating a cluster is given by,
\begin{equation}
\text{Ex}_{4,1}  = \frac{4nm}{3} - \frac{4n}{3} 
\end{equation}
In this notation, the `Ex' refers to the fact we are looking at the extra operations above the base line, while the subscript `$4,1$' refers to the fact that we are stitching together layers of width 4, with lines of qubits (i.e.~a layer of width 1). We use similar notation below, when we are considering the total number of operations needed to build the cluster, including the base $2nm$, then we use the notation `Cl'. If we use a subscript `layer' then the total number of operations is valid for several of our layered techniques (this will be made clear in the text). 

An alternative method involves using a path of width 4 for every other layer, with sealed boxes of width 2 providing the intermediate layers. This is illustrated in figure \ref{Fig:Gen:4box2}. In this case we need to repeat two rows of qubits (requiring an additional $2n$ operations per row) every $(m-4)/4$ rows. We also need to close the top and bottom using an additional $4n$ operations. Therefore in this case the total number of operations above our basic $2nm$ is given by
\begin{equation}
\text{Ex}_{4,2} = 4n(m-4)/4 +4n = nm \,.
\end{equation}
We now consider making a small adaption to this second technique. By having our cluster start and finish with a layer of width 2, we can increase the width of our cluster by 2 for an additional $4n$ operations. To illustrate how this changes the efficiency of the generation process, we replace $m$ with $m'$ where $m' = m+2$. The additional operations required to generate the extra layer are incorporated into our $2nm'$ therefore we only require
\begin{equation}
\text{Ex}'_{2,4} = \text{Ex}_{4,2} - 2nm + 2nm' = nm' - 2n
\end{equation}
above our basic number of $2mn$. Using layers of width 2 instead of simply stitching the ends is therefore slightly more efficient. 

An alternative technique involves using a layer of width 3, since this will only require stitching on one end. The layer of width 3 consists of a sealed layer of width 2 with free edges. If our entire cluster is made of layers of width 3, we see that $n(m-3)/2$ qubits interact with the bus as part of two separate sections. Each qubit that is part of 2 sections will require 2 operations above the basic number needed to make a cluster. An additional $2n$ operations are also required to seal the cluster, therefore a total of 
\begin{equation}
\text{Ex}_{3,3} = 2n(m-3)/2 +2n = nm - n
\end{equation}
extra operations are required to generate the entire cluster. Once again, we want to consider what happens if we use a layer of width 2 to close our cluster rather than just stitching it closed with a line of qubits. In this case we need to consider replacing m with $m' = m+1$. The additional operations needed to add this additional line of qubits are entirely contained within the $2nm'$ basic for a cluster. Therefore, the additional operations above the basic is given by
\begin{equation}
\text{Ex}'_{3,3} = nm' - 2n \,.
\end{equation}
We can see that we have two equally efficient techniques for generating the cluster. However, we don't necessarily need to disconnect the qubits completely from the bus between layers. If we leave some qubits entangled with the bus between layers, it is possible to further reduce the number of operations required to generate the cluster. 

\section{Saving operations when combining layers} \label{Comb}
\subsection{Notation for our diagrams} \label{notation} 
We will now establish a common notation for all future diagams. To reduce the total number of diagrams, we will display several operations in each. Each set of operations will consist of one sequence of operations where qubits disconnect from the bus, and one sequence of operations where qubits entangle to the bus. These operations often need to be performed in a specific order; the diagrams do not allow this order to be reconstructed simply by inspection, but they do clearly illustrate that such a sequence can be found. Once again, we will colour our qubits in diagrams where they are entangled to a particular quadrature of the bus: pink qubits are connected to the momentum quadrature, and blue qubits to the position quadrature. A qubit that is not entangled to either quadrature of the bus will be represented in black. 

We now need to consider how operations work together to make a gate. If we consider just two displacement operators then we have, 
\begin{equation}
D  \left ( \beta_{1}\sigma_{z1} \right )D \left (-i\beta_{2}\sigma_{z2} \right ) = \exp \left (i\beta_{1}\beta_{2}\sigma_{z1}\sigma_{z2} \right ) D\left (\beta_{1}\sigma_{z1}-i\beta_{2}\sigma_{z2} \right ) \,.
\end{equation}
This only gives us half of our desired interaction, $\exp(i\pi\sigma_{z1}\sigma_{z2}/4)$, and also results in a net displacement operation on the bus. We generate the second half of our interaction by removing the qubits from the bus such that
\begin{equation}
D \left (-\beta_{1}\sigma_{z1} \right)D \left(i\beta_{2}\sigma_{z2}\right) = \exp \left(i\beta_{1}\beta_{2}\sigma_{z1}\sigma_{2}\right) D\left(-\beta_{1}\sigma_{z1}+i\beta_{2}\sigma_{z2} \right)
\end{equation}
When combined we find that 
\begin{equation}
D\left (\beta_{1}\sigma_{z1}-i\beta_{2}\sigma_{z2} \right )D\left(-\beta_{1}\sigma_{z1}+i\beta_{2}\sigma_{z2} \right) = \exp(0) = 1
\end{equation}
so that
\begin{equation} 
D  \left ( \beta_{1}\sigma_{z1} \right )D \left (-i\beta_{2}\sigma_{z2} \right )D  \left ( -\beta_{1}\sigma_{z1} \right )D \left (i\beta_{2}\sigma_{z2} \right ) = \exp \left (2i\beta_{1}\beta_{2}\sigma_{z1}\sigma_{z2} \right ) \,.
\end{equation}
Therefore in our diagrams we want to show this interaction being created in two parts. A single line will represent the interaction $\exp(i\beta_{a}\beta_{b}\sigma_{za}\sigma_{zb})$ between qubits $a$ and $b$, with two lines being required for a maximally entangling interaction. Since all our interactions commute, these two lines do not need to be generated by sequential operations. An light green line will represent the fact that the interaction is being generated by displacement operators in the set illustrated on the diagram. A dark blue line will represent an interaction that has been generated previously. 

\begin{figure}[tb]
\centering
\subfigure[]
{\includegraphics[width=6cm]{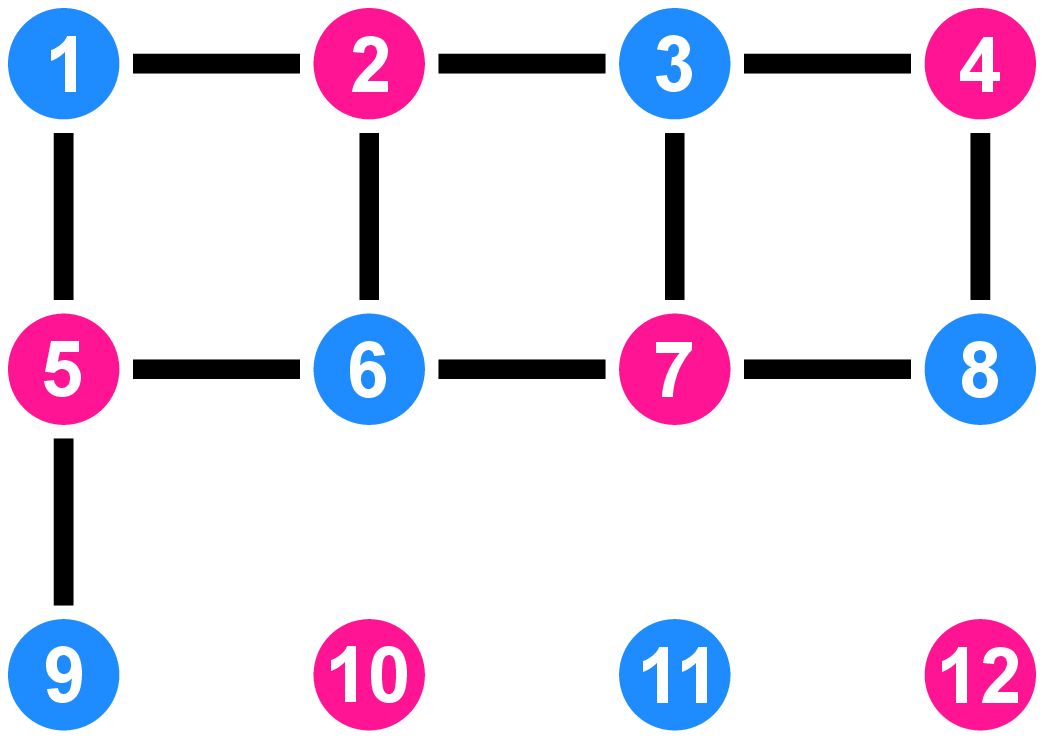}
\label{Fig:Layer:Tail2}}
\subfigure[]
{\includegraphics[width=6cm]{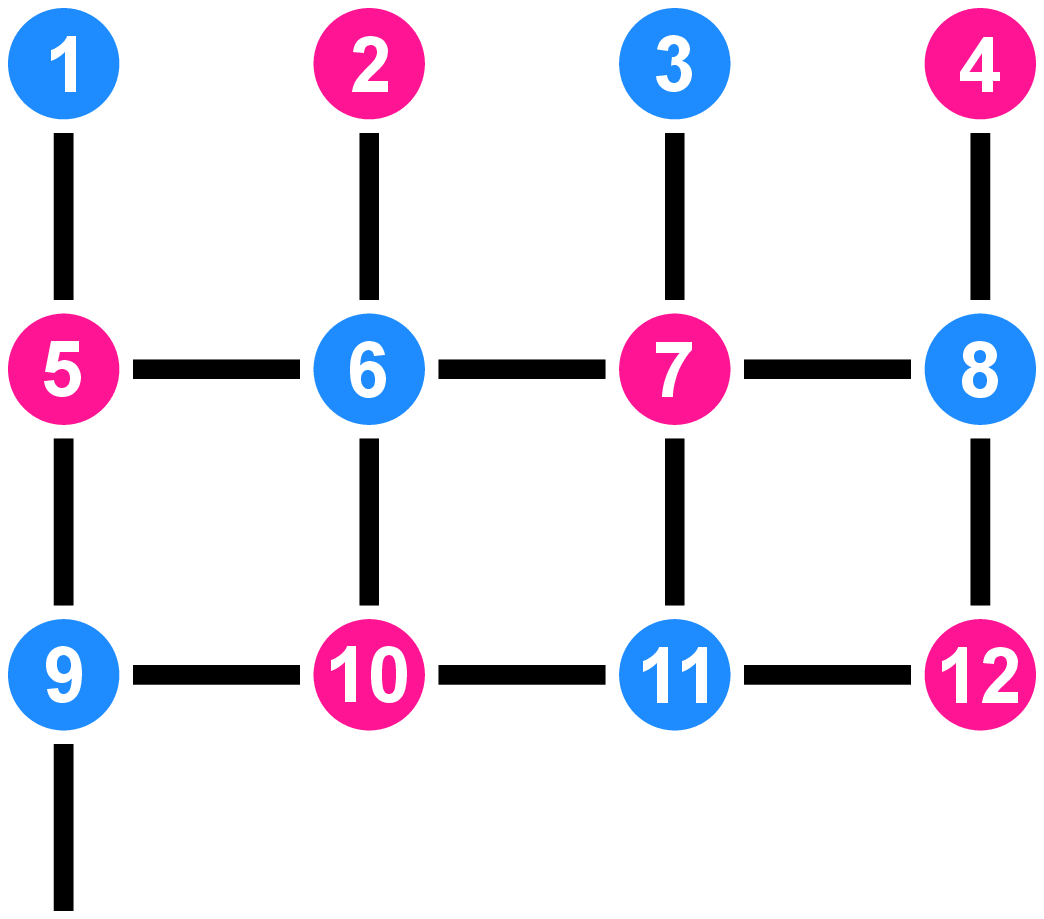}
\label{Fig:Layer:Tail3}}
\caption{(a) shows a layer of width 2 with a hanging edge. (b) shows a layer of width 3 with a hanging edge.} \label{Fig:Tail}
\end{figure}
It is possible to save at least two operations (a set of operations on a single qubit) per transition between layers since when starting a new layer we never have to deactivate the bottom corner qubit from the layer before. We can therefore see that it will always be possible to generate our cluster in 
\begin{equation}
\text{Cl}_{\text{layer}} = 3nm - 2n - m + 2 
\end{equation}
operations. However, in many cases it is possible to save an additional two operations by carefully choosing how we construct the connections between the layers, as we now describe.

We now need to introduce the concept of a `hanging edge'. This is a single extra C-Phase link created at the beginning of a layer, that hangs down into the next layer, as illustrated in figure \ref{Fig:Tail}.  If our hanging edge is connected to the first layer, then it has the potential to reduce the number of operations required to connect the second and third layer. We note that, if we are starting a layer of width 2, or a layer of width 3 with only the top corner qubit (qubit 1 in figure \ref{Fig:Tail}) connected to the bus, then it is always possible to create this hanging edge. This is unsurprising, since in this case we could go on and create a layer of width 4; therefore the hanging edge doesn't require any additional qubits to be removed from the bus for its creation. 

It is important to note that the addition of a hanging edge doesn't require any extra operations in most cases. In figure \ref{Fig:Layer:Tail2} two extra operations are required to create the layer illustrated; however these are saved when we create the next layer. The addition of the hanging edge simply substitutes interacting qubit 5 with the bus four times, with interacting qubit 9 with the bus four times. In the next subsections, we will discuss exactly when it is possible to save operations because this hanging edge has been created, and why this is the case. 

\subsection{Transferring between a layer of width 2 and a layer of width 3. }

\begin{figure}[tb]
\centering
\subfigure[]
{\includegraphics[width=4.6cm]{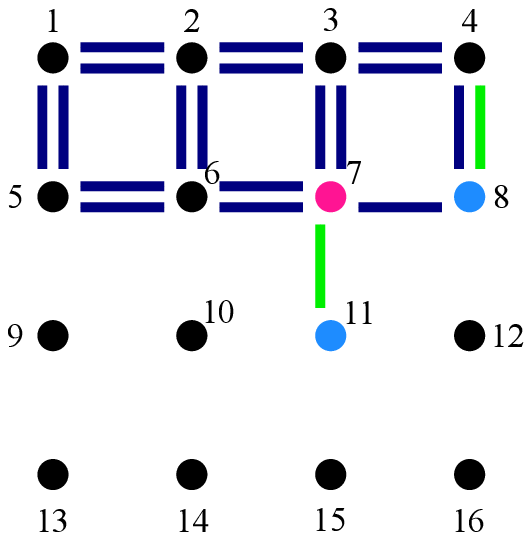}
\label{Fig:Gen:transition237}}
\subfigure[]
{\includegraphics[width=4.6cm]{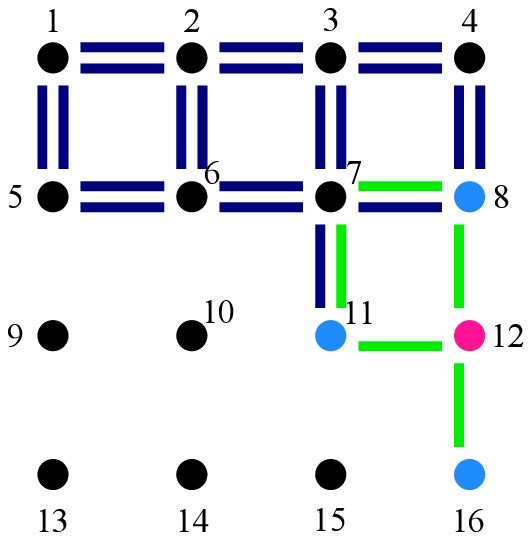}
\label{Fig:Gen:transition238}}
\subfigure[]
{\includegraphics[width=4.6cm]{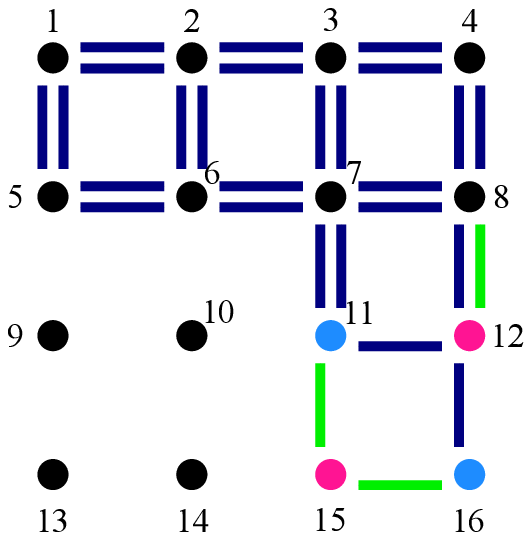}
\label{Fig:Gen:transition239}}
\caption{Transferring between a layer of width 2 and a layer of width 3. (a) shows that we can keep qubits 7 and 8 entangled to the bus after the layer of width 2. (b) and (c) show that it is then possible to create a layer of width 3. The colour coding of the diagram is described in Section \ref{notation}.} \label{Fig:Gen:transition23}
\end{figure}

Figure \ref{Fig:Gen:transition23} shows that it is possible to connect our layer of width 2 to a layer of width 3, while saving 4 operations compared to generating each layer separately. It is worth noting that we would not be able to add a hanging edge to this diagram, without first disconnecting one of the qubits from the bus then reconnecting it. This is because qubit 11 needs to connect to the position quadrature of the bus before we entangle qubit 16 to the position quadrature of the bus. This is because we want qubit 11 to entangle with qubit 7, which needs to connect to the bus at an earlier stage. However this means that, to generate our hanging edge, we need to remove qubit 11 from the bus, or we will generate additional unwanted C-Phase gates. As a consequence, we are limited in what we are able to perform, and in this case we can not create a hanging edge. 

\subsection{Transferring between two layers of width 3} \label{Sub:2layer3}

\begin{figure}[tb]
\centering
\subfigure[]
{\includegraphics[width=4.6cm]{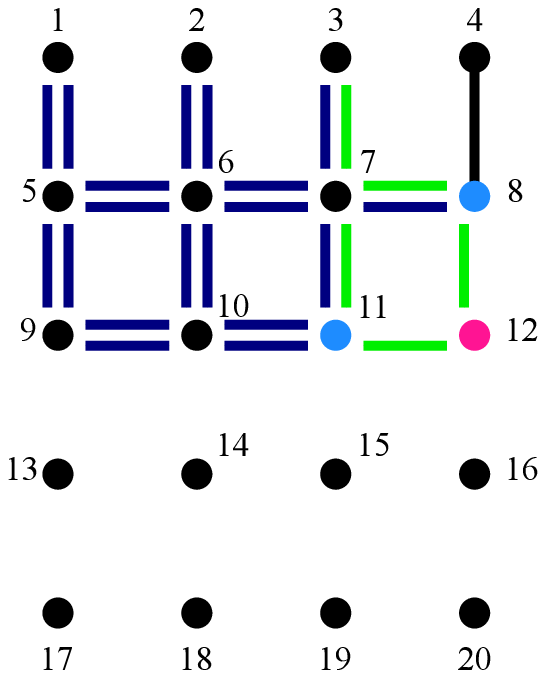}
\label{Fig:Gen:tail337}}
\subfigure[]
{\includegraphics[width=4.6cm]{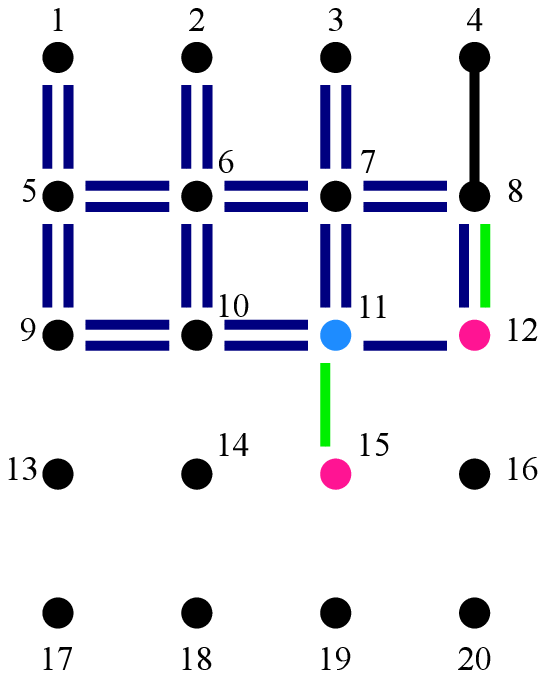}
\label{Fig:Gen:tail338}}
\subfigure[]
{\includegraphics[width=4.6cm]{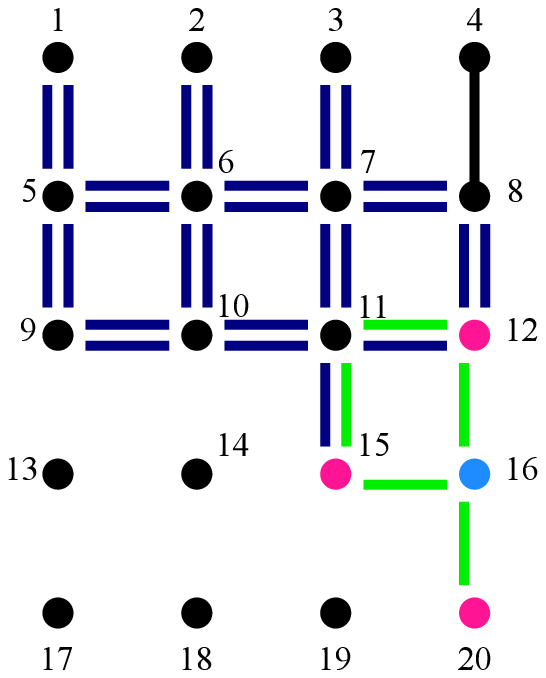}
\label{Fig:Gen:tail339}}
\caption{Transferring between two layers of width 3. (a) shows that if we have a hanging edge between qubits 4 and 8 we can create the first layer of width 3 while leaving qubits 11 and 12 entangled to the bus. (b) and (c) show that we can then build another layer of width 3 without reattaching qubits 11 and 12 to the bus. The colour coding of the diagram is described in Section \ref{notation}.} \label{Fig:Gen:tail33}
\end{figure}

In figure \ref{Fig:Gen:tail33} we can see that, if we have the hanging edge shown in black between qubits 4 and 8 already created, then we can make a saving of four operations when transferring between two layers of width 3. Without this hanging edge, only two operations could be saved. We can see this by looking at figure \ref{Fig:Gen:tail337}, where we would need to remove and then reconnect qubit 11 from the bus so that we could create the operation shown in black. The hanging edge provides us with a significant saving in operations since it removes the needs for qubits 8 and 16 to both connect to separate qubits at the same time. Both of these qubits connect to qubit 12, as well as their corner qubits, 4 and 20. With the limitations of ordering, this means we can not create the extra links without disconnecting and then reconnecting qubit 11, or an alternative nearby qubit.

\subsection{Transferring between a layer of width 4 and a layer of width 2}

\begin{figure}[tb]
\centering
\subfigure[]
{\includegraphics[width=4.6cm]{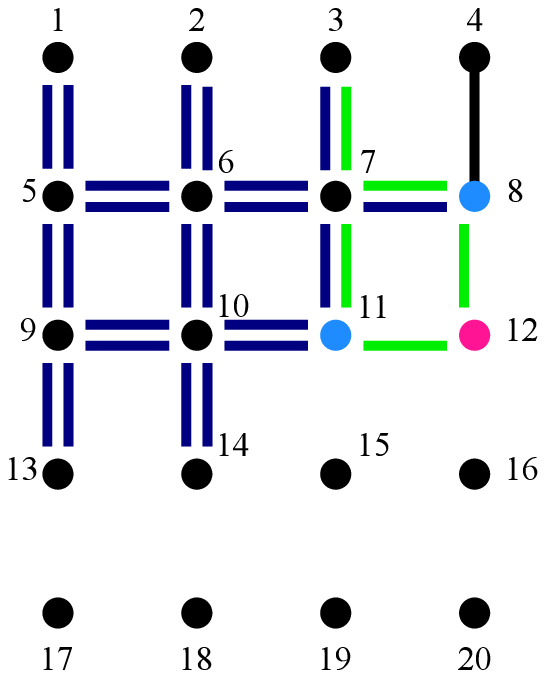}
\label{Fig:Gen:tail422}}
\subfigure[]
{\includegraphics[width=4.6cm]{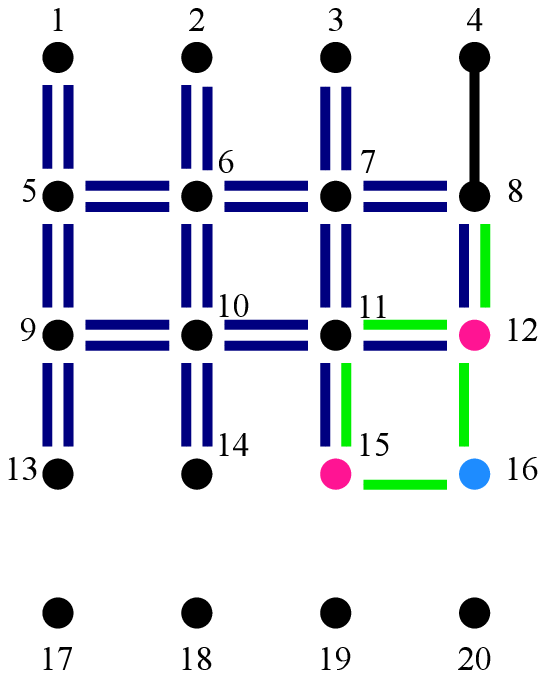}
\label{Fig:Gen:tail423}}
\subfigure[]
{\includegraphics[width=4.6cm]{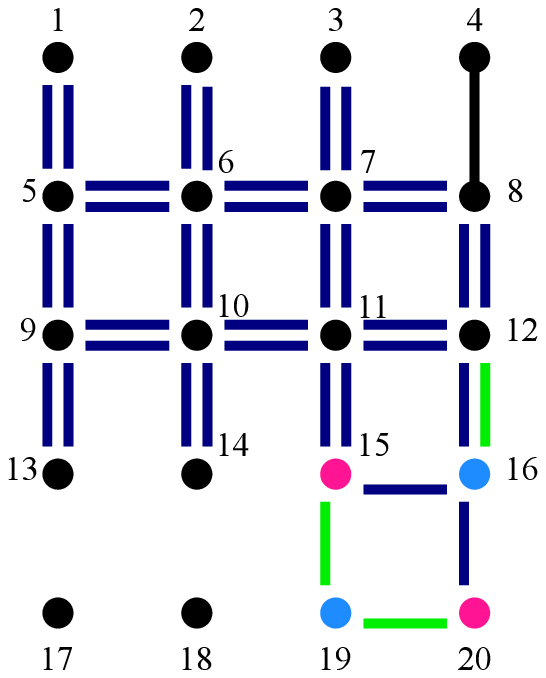}
\label{Fig:Gen:tail425}}
\caption{Transferring between a layer of width 4, and a layer of width 2. (a) and (b) show how to build a layer of width 4 leaving qubits 15 and 16 entangled with the bus. (c) shows that it is then possible to create a layer of width 2. The colour coding of the diagram is described in Section \ref{notation}.} \label{Fig:Gen:tail42}
\end{figure}

We also want to consider the transitions when we have layers that alternate between being of width 2 and width 4. Figure \ref{Fig:Gen:tail42} demonstrates that it is possible to save 4 operations when transitioning from a layer of width 4 to a layer of width 2, provided that we have the hanging edge shown in black already generated. As before, it is impossible to save these 4 operations if we do not have this hanging edge generated, since we would need to reactivate qubit 11. This is because qubit 15 and qubit 20 both need to connect to qubit 16, as well as separate qubits (11 and 19 respectively). As we do not want to remove qubit 15 from the bus, but have to connect it, either before or within the same sequence of operations as qubit 20, we would need to remove one of the nearby qubits from the bus to create all the necessary operations. However, if the edge shown in black weren't already in place, then it would be possible to create an additional hanging edge: since we have to disconnect then reconnect qubit 11 in this case, we can connect qubit 15 at a later stage than in the case where we don't have to do this.  

\subsection{Transferring between a layer of width 2 and a layer of width 4}

If we look at a transition between a layer of width 2, and a layer of width 4, then a pre-created hanging edge will not provide any saving in the number of operations required. This is because the extra operations only help create the layer of width 2, something we can do without too much difficulty. The hanging edge will not change the order in which we perform the operations, and it is this ordering which limits our ability to create the necessary layer of width 4. 

\section{Total number of operations required using this layered technique} \label{Tot}
\begin{table}[tb]
\centering
\begin{tabular}{|c|c|c|c|c|}
\hline
Size of layers & Hanging edge available? & Saving in transfer & Generate hanging edge? \\
\hline \hline
$2 \rightarrow 3$ & Not important & 4 & No \\
\hline \hline
$3 \rightarrow 3$ & No & 2 & Yes \\
\hline
$3 \rightarrow 3$ & Yes & 4 & No \\
\hline \hline
$2 \rightarrow 4$ & Not important & 2 & Yes \\
\hline \hline
$4 \rightarrow 2$ & No & 2 & Yes \\
\hline
$4 \rightarrow 2$ & Yes & 4 & No \\
\hline 
\end{tabular}
\caption[Savings available between layers]{The savings that can be obtained when generating the transition between layers of the cluster.} \label{Tab:Layer:Saving}
\end{table}

We can use the information about how many operations we can save per transition to give the total number of operations required to generate a cluster that has a large value of $m$, and which can be exactly divided into these layers. Table \ref{Tab:Layer:Saving} summarises the results discussed in this section. A key point is that a saving of 4 operations prevents us from generating a hanging edge, and thereby achieving a saving in the transition after next. Therefore, our results are simplest for the case where our cluster is made of layers of width 3. Here we will see a pattern of two transitions where we save 4 operations, two transitions where we save 2 operations, two transitions where we save 4 operations and so on. For a large cluster the number of transitions is given by $(m-2)/2$, and therefore our total saving is given by $3(m-2)/2$. This gives 
\begin{equation}
\text{Cl}_{3,3} = 3nm - 2n -\frac{3}{2}m + 3 \,.
\end{equation}
We can see that this is $(m-2)/2$ operations above the lower bound given by 
\begin{equation}
\text{Cl}_{\text{min}} = 3nm - 2n - 2m + 4
\end{equation}
which we showed in previous work \cite{Horsman2010}.

The savings we find in the case where we use our layers of width 4 are less significant. In this case we save 2 operations for 3 out of every 4 transitions, only saving 4 operations for a quarter of our transitions. The total saving is therefore given by $5(m-2)/4$, therefore
\begin{equation}
\text{Cl}_{2,4} = 3nm - 2n - \frac{5}{4}m + \frac{5}{2} \,. 
\end{equation}
This is $3(m-2)/4$ operations above our lower bound -- significantly more operations than needed to create our cluster using layers of width 3. 

\section{A layered technique using negative operations} \label{Neg}

We now consider what is possible when we can create and then destroy links. To do this, we add in some additional notation. If a dark blue line disappears in a figure, this means we are generating the interaction $\exp(-i\beta_{a}\beta_{b}\sigma_{za}\sigma_{zb})$, and thus destroying the previously generated interaction. In the qubus architecture gates are created provided we sandwich the operations on one quadrature of the bus with the operations on the other. For example, the sequence of displacement operators 
\begin{equation}
U_{12} = D(\beta_{1}\sigma_{z1})D(-i\beta_{2}\sigma_{z2})D(\beta_{1}\sigma_{z1})D(i\beta_{2}\sigma_{z2})
\end{equation}
creates a net gate given by 
\begin{equation}
U_{12} = \exp(2i\beta_{1}\beta_{2}\sigma_{z1}\sigma_{z2}) \,.
\end{equation} 
However an unsandwiched set of operations such as 
\begin{equation}
U_{0} = D(\beta_{1}\sigma_{z1})D(-i\beta_{2}\sigma_{z2})D(i\beta_{2}\sigma_{z2})D(-\beta_{1}\sigma_{z1})
\end{equation}
leads to no net operation on either the qubits or the bus. We can therefore imagine a scenario where we have qubit 1 and qubit 2 connected to the momentum quadrature of the bus, then we connect qubit 3 to the position quadrature. If we remove qubit 1, then qubit 3, then finally qubit 2, we can create entanglement between qubit 1 and qubit 3, while leaving qubit 2 completely disentangled from the system. This is shown below 
\begin{equation}
\begin{split}
U_{13} &= D(\beta_{1}\sigma_{z1})D(\beta_{2}\sigma_{z2})D(-i\beta_{3}\sigma_{z3})D(-\beta_{1}\sigma_{z1})D(i\beta_{3}\sigma_{z3})D(-\beta_{2}\sigma_{z2}) \\
& = \exp(2i\beta_{1}\beta_{3}\sigma_{z1}\sigma_{z3})
\end{split}
\end{equation}

When we use these extra links in our cluster state generation they do not require extra operations (unlike in the simple example above). They are created by leaving particular qubits on the bus longer than we did previously. These qubits them interact with subsequent qubits as they are added and removed from the bus, only being removed when any extra links which have been created have already been destroyed. Using this technique, it is possible to leave two qubits entangled to the bus for each transition between layers. We therefore save a total of four operations. In section \ref{Simple} we established that we switch layers $(m-3)/2$ times; however this excluded the addition of the extra layers of width 2. Once we include these, we find that we switch layers a total of $(m-2)/2$ times. This means we can save a total of $2(m-2)$ operations as part of our layer changes, bringing the total number of operations required down to 
\begin{equation}
\text{Cl} = 3nm - 2n -2m +4
\end{equation}
This is the same number of operations as the lower bound we derived in previous work \cite{Horsman2010}, although we have now gone beyond the assumptions of the previous paper. 
\begin{figure}[tb]
\subfigure[]
{\includegraphics[width=4.8cm]{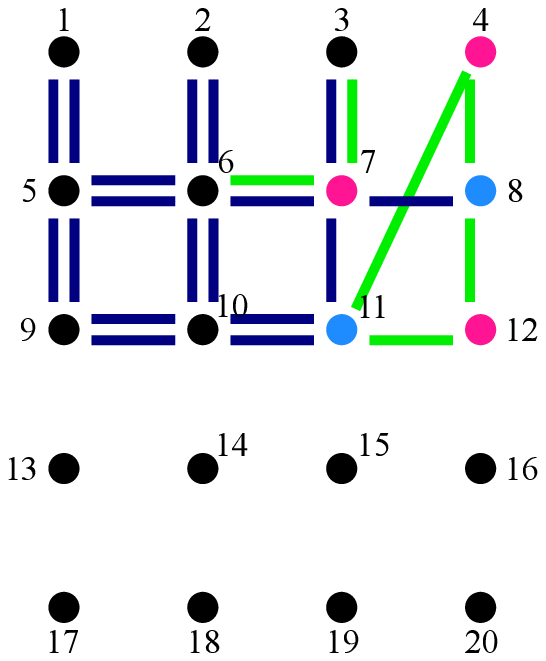}
\label{Reverse3}}
\subfigure[]
{\includegraphics[width=4.8cm]{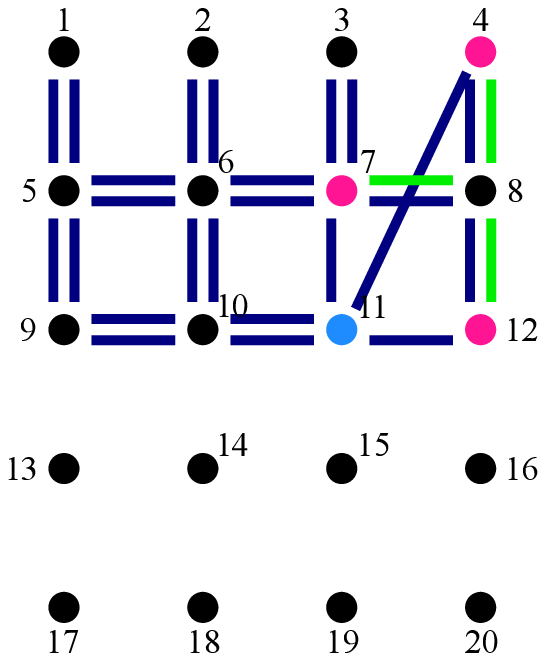}
\label{Reverse4}}
\subfigure[]
{\includegraphics[width=4.8cm]{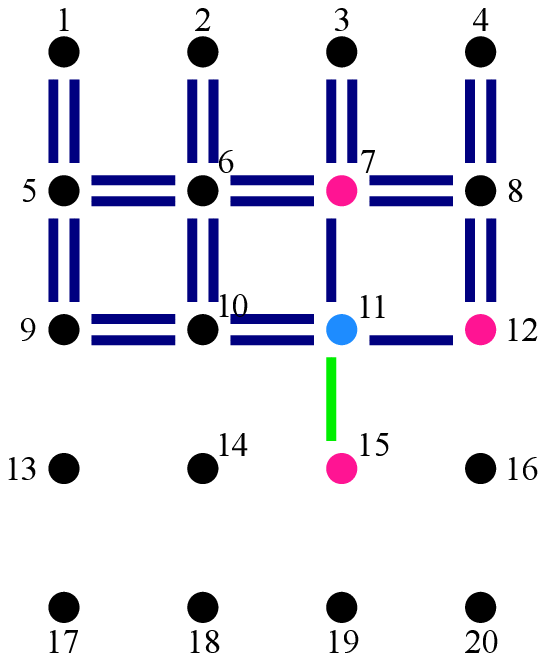}
\label{Reverse5}}
\caption{Saving operations by creating then destroying links. (a) shows the creation of an unwanted operation between qubits 4 and 11. (b) shows this operation being destroyed. (c) shows we can then create the next layer without reactivating any other qubits. The colour coding of the diagram is described in Section \ref{notation}.} \label{Reverse} 
\end{figure}

To demonstrate that we can save this number of operations, we need to look at the transition between two layers of width 3. The connections between a layer of width 2 and a layer of width 4, and a layer of width 4 and a layer of width 2 work similarly. We do not illustrate these here, because the techniques are similar, and a proof that we can save four operations in this case is not necessary to show we can saturate our lower bound. Figure \ref{Reverse} illustrates line 2 and 3 of equation (\ref{Requation}). These are the most significant steps to illustrate since they show how we create the corner, which would be impossible without creating, then destroying the additional diagonal interaction in figure \ref{Reverse3}.   

\begin{equation}
\begin{split}
&U_{L1} = D(\beta_{1}\sigma_{z1})D(\beta_{9}\sigma_{z9})D(-i\beta_{5}\sigma_{z5})D(-\beta_{1}\sigma_{z1})D(-i\beta_{10}\sigma_{z10})D(-\beta_{9}\sigma_{z9})D(-i\beta_{2}\sigma_{z2})D(-\beta_{6}\sigma_{z6}) \\
&D(i\beta_{5}\sigma_{z5})D(i\beta_{2}\sigma_{z2})D(-\beta_{11}\sigma_{z11})D(i\beta_{10}\sigma_{z10})D(-\beta_{3}\sigma_{z3})D(-\beta_{8}\sigma_{z8})D(i\beta_{7}\sigma_{z7})D(\beta_{6}\sigma_{z6})D(\beta_{3}\sigma_{z3})\\
&D(i\beta_{4}\sigma_{z4})D(i\beta_{12}\sigma_{z12})D(\beta_{8}\sigma_{z8})D(-i\beta\sigma_{z4})D(i\beta_{15}\sigma_{z15})D(\beta_{11}\sigma_{z11})D(-i\beta_{7}\sigma_{z7})D(i\beta_{20}\sigma_{z20})D(\beta_{16}\sigma_{z16}) \\
&D(-i\beta_{12}\sigma_{z12})D(\beta_{19}\sigma_{z19})D(-i\beta_{20}\sigma_{z20})D(\beta_{14}\sigma_{z14})D(-i\beta_{15}\sigma_{z15})D(-\beta_{16}\sigma_{z15})D(-i\beta_{18})\sigma_{z18})D(-i\beta_{10}\sigma_{z10}) \\ 
&D(-i\beta_{13}\sigma_{z13})D(-\beta_{14}\sigma_{z14})D(i\beta_{10}\sigma_{z10})D(-\beta_{17}\sigma_{z17})D(i\beta_{18}\sigma_{z18})D(-\beta_{9}\sigma_{z9})D(i\beta_{13}\sigma_{z13})D(\beta_{17}\sigma_{z17})D(\beta_{9}\sigma_{z9})
\end{split} \label{Requation} 
\end{equation}

\section{Revisiting the limitations of the bus} \label{Rev}
In section \ref{Sec:Limitations} we established that if we don't use `negative' operations, then the largest layer we can create has width 4. However, now we have introduced the idea of negative operations this is no longer valid. Similarly while we demonstrate that we can meet the lower bound we derived in previous work \cite{Horsman2010}, this lower bound also made the assumption that we couldn't destroy previously created links. We therefore want to see if it is possible to create a larger layer while only interacting each qubit with the bus twice. We will start by showing it is possible to create a $3\times3$ box shown in figure \ref{Fig:Gen:box} using a sequence of displacement operators given by 
\begin{equation}
\begin{split}
U_{B} = & D(\beta_{z1}\sigma_{z1})D(-i\beta_{2}\sigma_{z2})D(-i\beta_{4}\sigma_{z4})D(-\beta_{1}\sigma_{z1})D(-i\beta_{z8}\sigma_{z8})D(-\beta_{5}\sigma_{z5})D(-\beta_{7}\sigma_{z7})D(i\beta_{4}\sigma_{z4})D(-\beta_{9}\sigma_{z9})\\
& D(i\beta_{8}\sigma_{z8})D(\beta_{7}\sigma_{z7})D(-\beta_{3}\sigma_{z3})D(i\beta_{6}\sigma_{z6})D(\beta_{9}\sigma_{z9})D(i\beta_{2}\sigma_{z2})D(\beta_{3}\sigma_{z3})D(\beta_{5}\sigma_{z5})D(-i\beta_{6}\sigma_{6}) \,.
\end{split}
\end{equation}
This sequence of operations leaves qubit 2 connected to the bus, creating and then destroying operations between it and qubit 7, and qubit 9. The result is a small cluster of $3\times3$ qubits without any additional operations. This allows us to increase the largest layer of the cluster we can build, and thus beat the lower bound. In figure \ref{sixwide} in appendix \ref{Appendix} we demonstrate that it is possible to create a layer of width 4, with one sealed end and one open end, while only interacting each qubit with the bus twice. While we don't show it is impossible to do the same thing for a layer of width 4 sealed on both ends, choosing the gate sequence in this case becomes more difficult, owing to the restrictions on when we remove certain qubits from the bus. If we include the necessary operations to seal our cluster, using a completely sealed layer of width 2, we reduce the number of transitions between layers down to $(m-2)/3$. By choosing our ordering carefully, we can make the transitions by reactivating $(n-2)$ qubits. This would require a total of $(8nm-4(n+m)/-8)/3$ to create the cluster, thus beating our lower bound. Saving any further operations when transitioning between layers becomes difficult and we do not show a technique to do this. 

It would be potentially possible to reduce this further by creating even larger layers, but this would require a more complex sequence of operations, and significantly reduce the dynamic nature of our scheme. Another significant difficulty would be that the time the qubits are left entangled with the bus would grow, increasing the probability of an error. Even with the layers of 4 with one sealed edge currently being used, if the length of the layer is odd, a qubit from each layer has to remain entangled with the bus until the entire cluster has been generated. This means that the cluster could not be used until the whole thing had been generated, a significant disadvantage compared to previous schemes. Interestingly, this isn't the case if the length of the cluster is even in which case the generation retains its dynamic nature. 

\section{Conclusions} \label{Conclusions}
In this paper, we have shown how to create a cluster state using the qubus quantum computer, using considerably fewer operations than previously required. In particular we show that, by combining operations, we can get over a factor of 4 savings over a na\"ive method, and a factor of 2 saving over the method presented by Louis et al.~\cite{Louis2007}. We consider three cases, starting by placing a restriction on the operations we can create, so we never create then destroy a link between two qubits. This prevents us from having unwanted interactions between qubits, if these operations don't cancel exactly. In this case we show how to create a $n\times m$ cluster state using only $3nm - 2n -\frac{3}{2}m + 3$ operations. 

However, we can reduce the total number of operations by introducing bus operations which create and then destroy unwanted interactions between qubits. The simplest method of doing this just gives us savings when we make a transition between layers. In a previous paper we derived a lower bound for a case where we didn't create these unwanted interactions \cite{Horsman2010}, and we now show that by adding them into the layered case we can meet this lower bound. We therefore provide a simple technique for creating a cluster using just $3nm - 2n - 2m +4$ operations. While this increases the risk of a net unwanted interaction (particularly if our operations are imperfect), the scheme is simple, and the interaction we need to create and destroy is not left to decohere for a long time before it is destroyed. 

By complicating our scheme further, we demonstrate how to significantly beat our lower bound, and generate an $n\times m$ cluster in only $\frac{1}{3}(8nm-4(n+m)/-8)$ operations. The disadvantage of this technique is the sheer complexity of the scheme. In particular, some operations sit on the bus creating unwanted interactions between themselves and a large proportion of the qubits in the cluster, and later destroying them. This significantly increases the chance of an error occurring, and one of these unwanted operations continuing to exist. Similarly while the other schemes are dynamic, in this case, the length of time an operation sits on the cluster increases as $m$ increases. Therefore the beginning of the cluster can't be used before the end of the cluster is created. This removes one of the principle advantages of the layered scheme. The first two methods of creating the cluster are therefore more practical for creating a large cluster, even though they are slightly less efficient. 

While we have discussed these results in terms of the qubus architecture in this paper, they are applicable to a wide range of ancilla based systems in any architecture with a bus that creates operations using two interactions. In previous work we have looked at how the errors propagate within such a scheme, and suggested that to prevent the probability of error increasing too much it would be necessary to split our cluster into smaller brick sections that can be built simultaneously \cite{Horsman2010}. By applying techniques from that work to the idealised results in this paper, we have techniques for generating clusters in the presence of realistic errors. 

\begin{acknowledgements}
KLB was supported by a UK EPSRC CASE studentship from Hewlett Packard. VMK is funded by a UK Royal Society University Research Fellowship. WJM was supported in part by MEXT and FIRST in Japan.
\end{acknowledgements}

\bibliography{bibliography}

\appendix
\section{Connecting two layers of width 4} \label{Appendix} 
\begin{figure}
\subfigure[]
{\includegraphics[width=4.8cm]{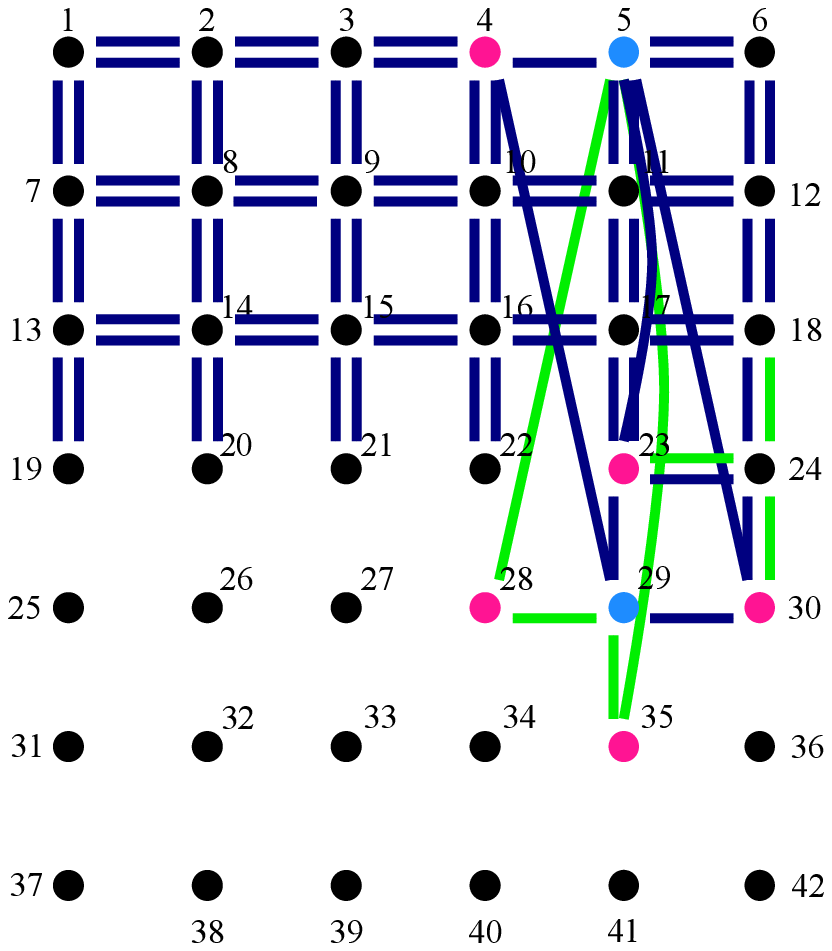}
\label{six13}}
\subfigure[]
{\includegraphics[width=4.8cm]{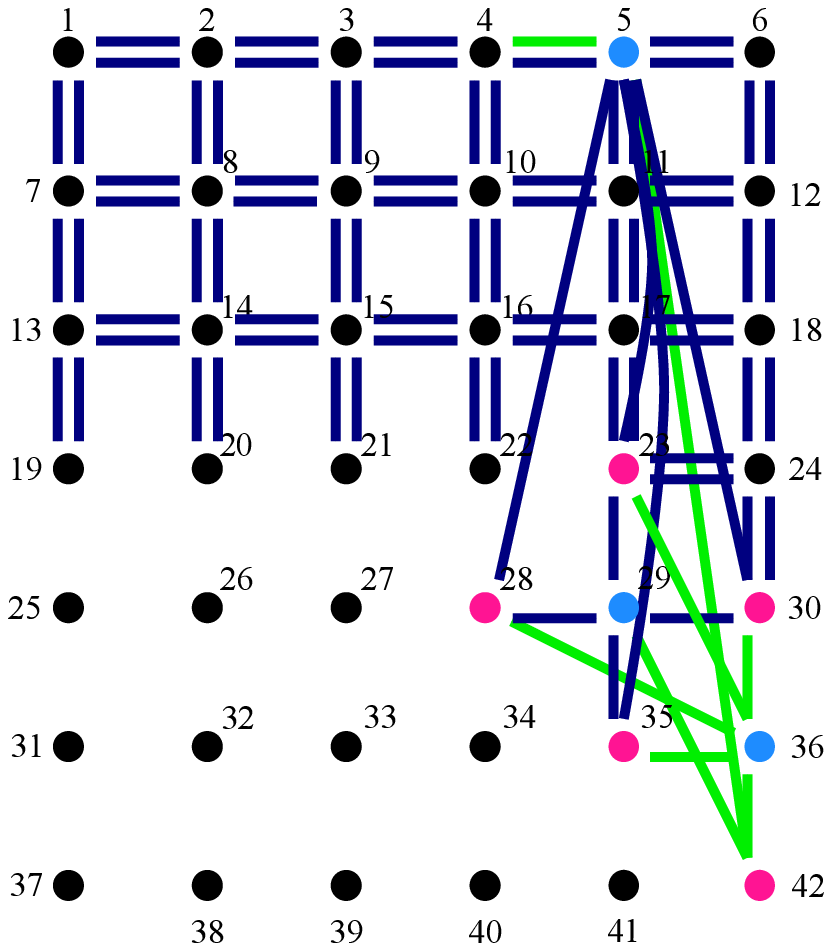}
\label{six14}}
\subfigure[]
{\includegraphics[width=4.8cm]{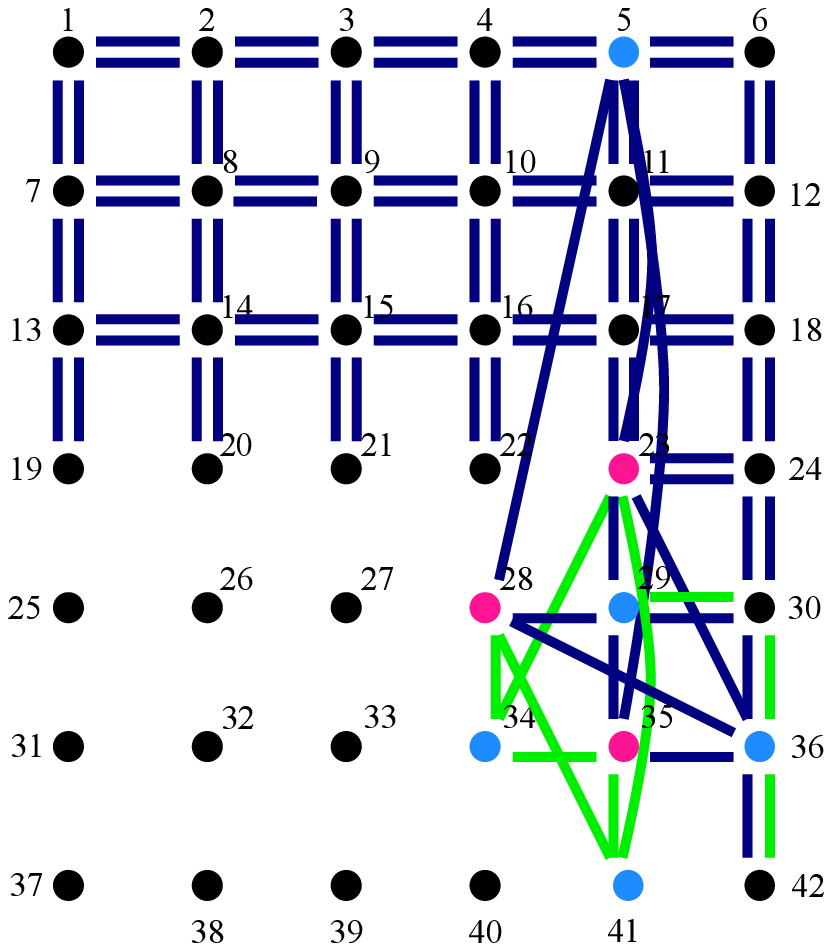}
\label{six15}}
\subfigure[]
{\includegraphics[width=4.8cm]{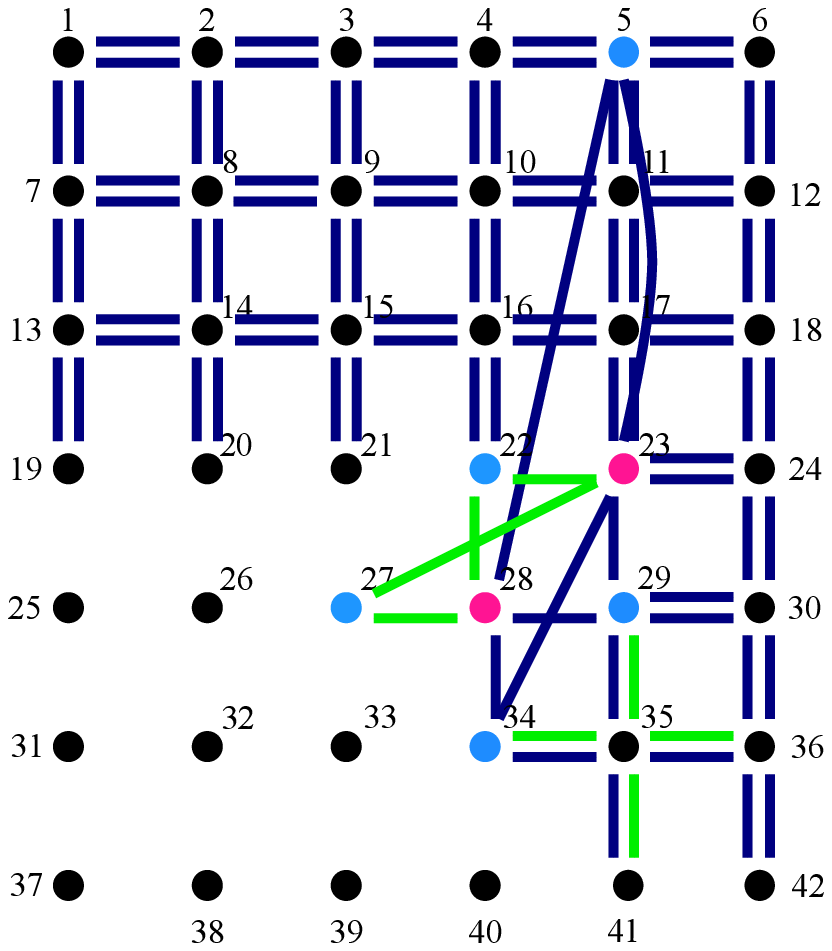}
\label{six16}}
\subfigure[]
{\includegraphics[width=4.8cm]{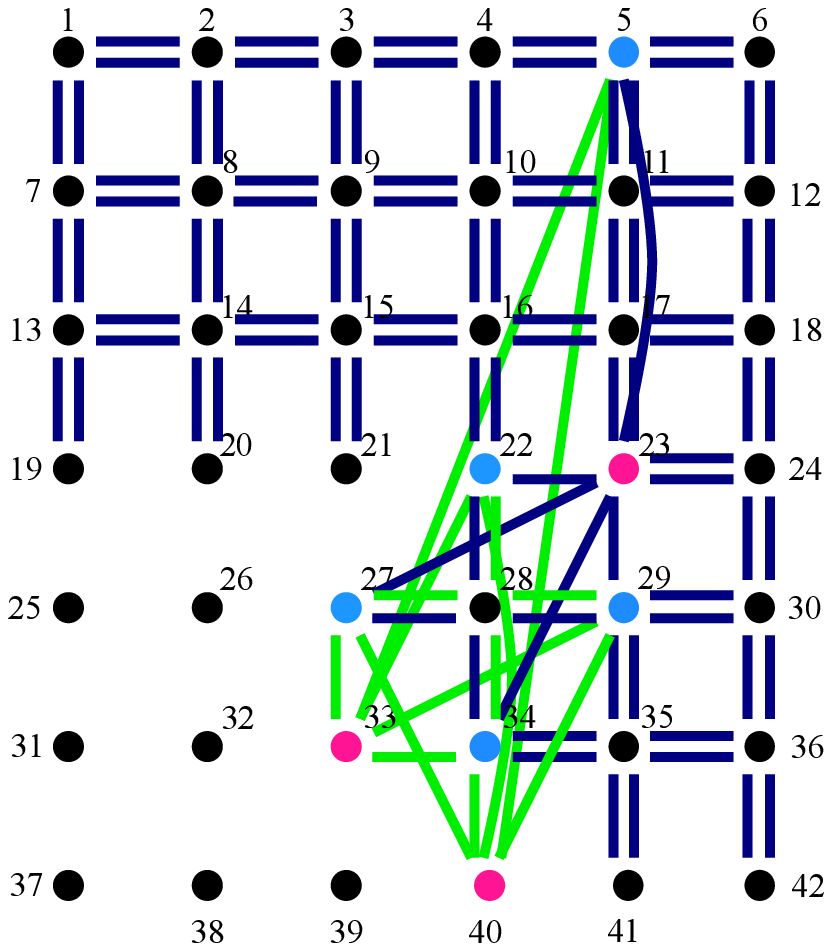}
\label{six17}}
\subfigure[]
{\includegraphics[width=4.8cm]{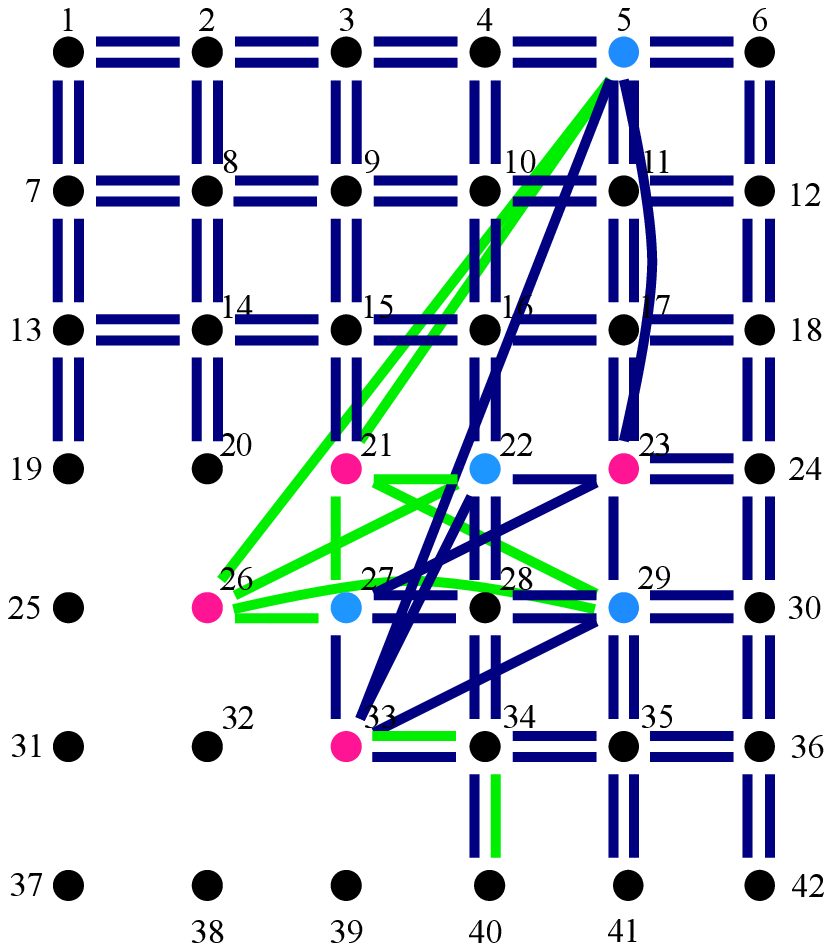}
\label{six18}}
\subfigure[]
{\includegraphics[width=4.8cm]{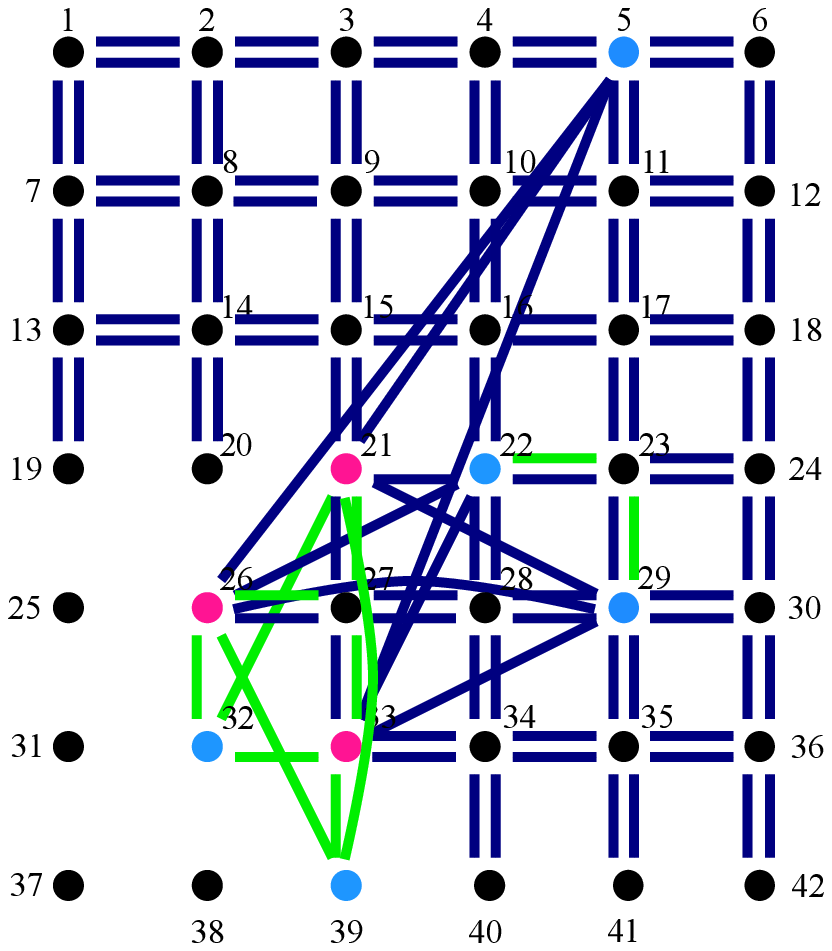}
\label{six19}}
\subfigure[]
{\includegraphics[width=4.8cm]{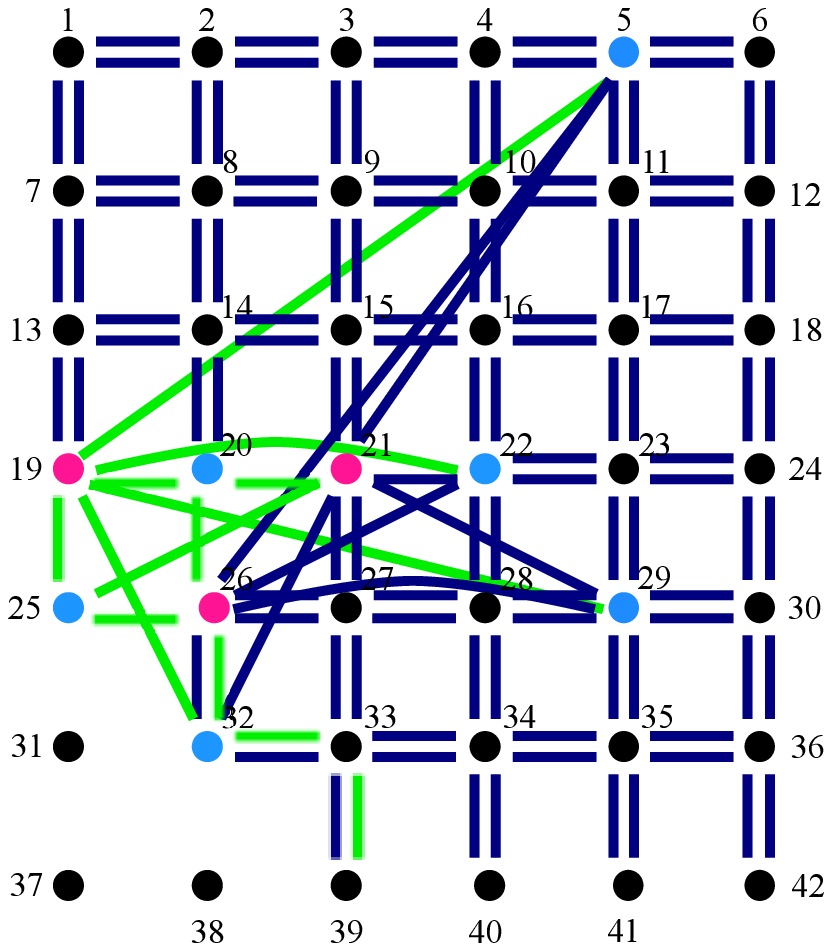}
\label{six20}}
\subfigure[]
{\includegraphics[width=4.8cm]{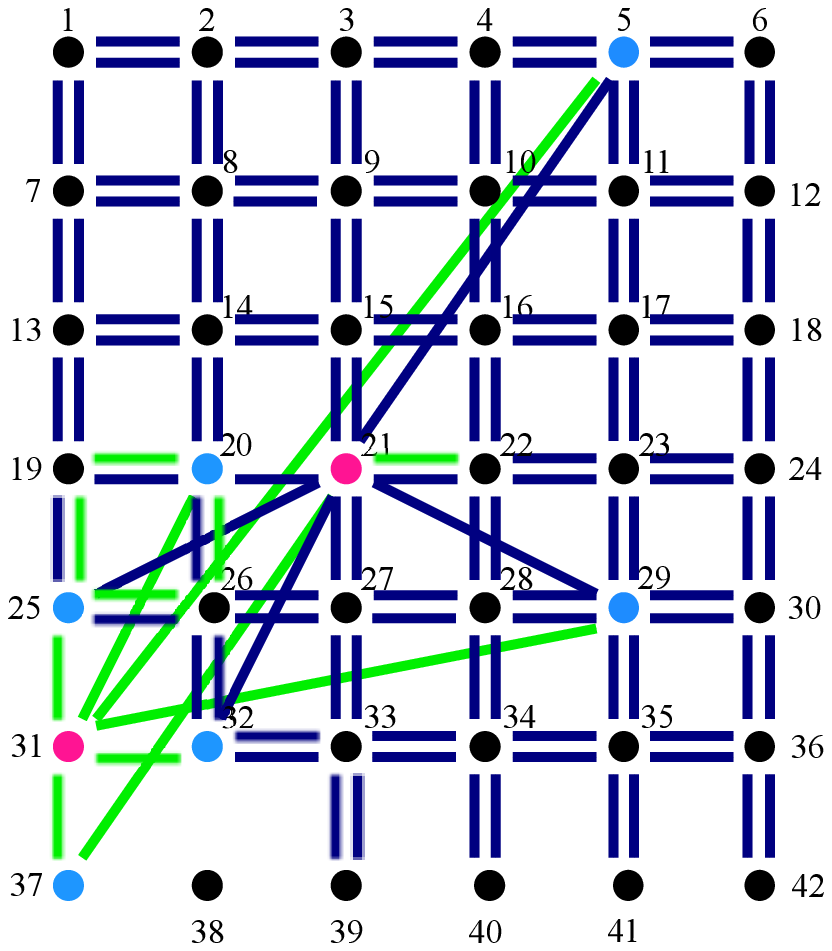}
\label{six21}}
\subfigure[]
{\includegraphics[width=4.8cm]{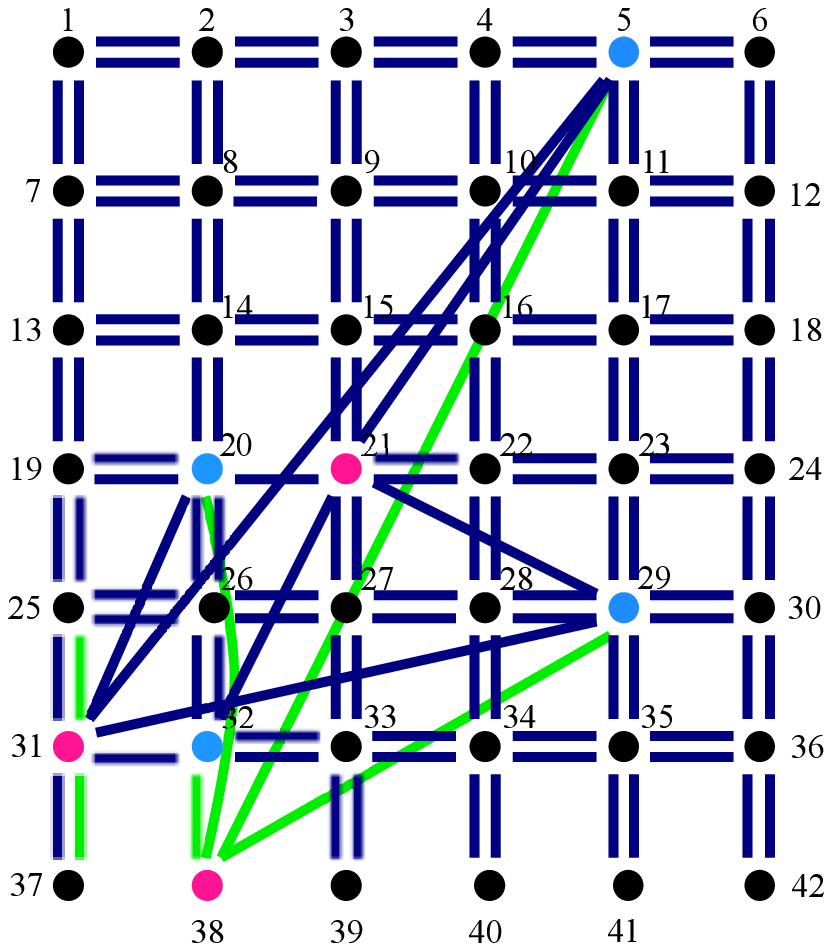}
\label{six22}}
\subfigure[]
{\includegraphics[width=4.8cm]{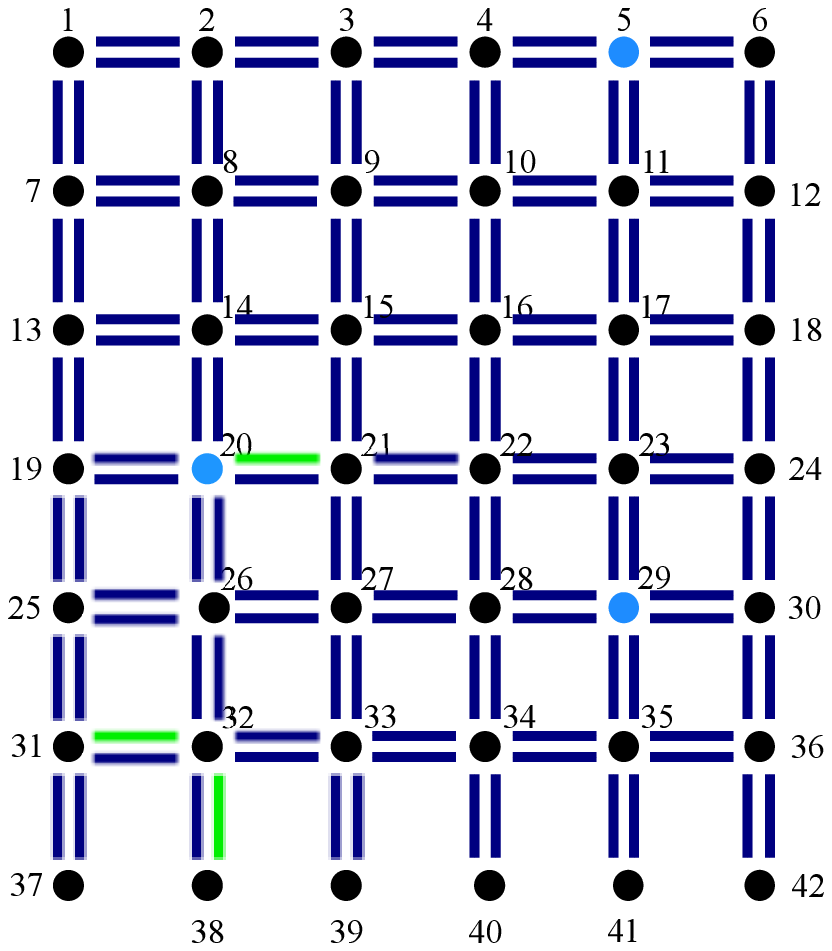}
\label{six23}}
\subfigure[]
{\includegraphics[width=4.8cm]{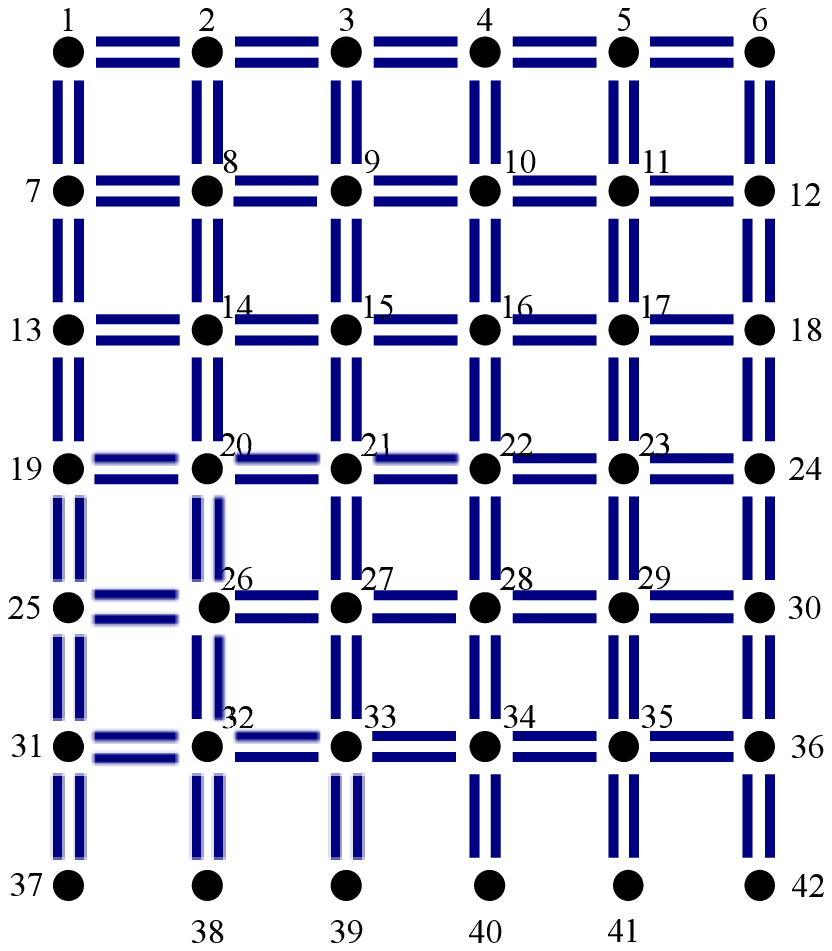}
\label{six24}}
\caption{Connecting two layers of width 4 while saving two operations. (a)-(i) show how a layer of width 4 can be created by leaving qubits on the bus. (j)-(n) show the transition between the two layers while leaving two qubits entangled to the bus. (o)-(x) show the generation of the second layer of width 4. The colour coding of the diagram is described in Section \ref{notation}.}
\label{sixwide}
\end{figure}

It is possible to connect two layers of width 4 which are closed on one end, and open on the other together while leaving two qubits entangled to the bus. Figure \ref{sixwide} illustrates this process. The key thing to note is that unwanted interactions are constantly being created and destroyed. The corner qubits, 23 and 24 only interact with the bus twice. Therefore we save 4 operations when transitioning between our layers. Each layer can be built only interacting each qubit in the layer with the bus twice. We can therefore clearly see that creating these interactions which are later destroyed does not cost any extra operations. 
\begin{figure}[h,p]
\subfigure[]
{\includegraphics[width=4.8cm]{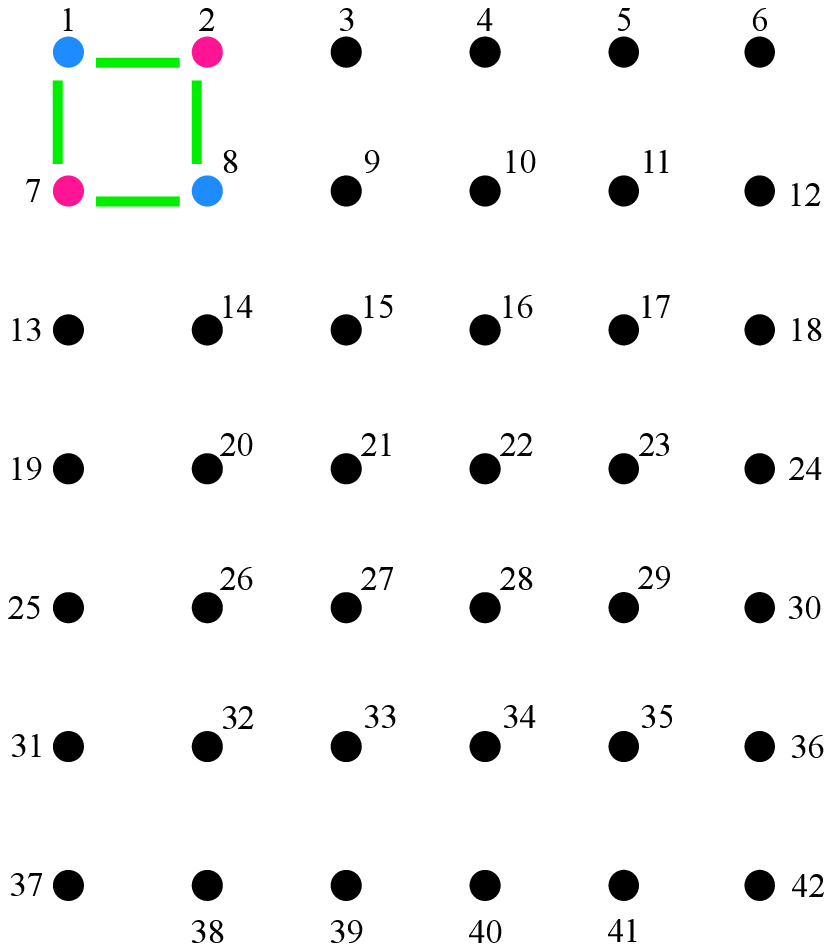}
\label{six1}}
\subfigure[]
{\includegraphics[width=4.8cm]{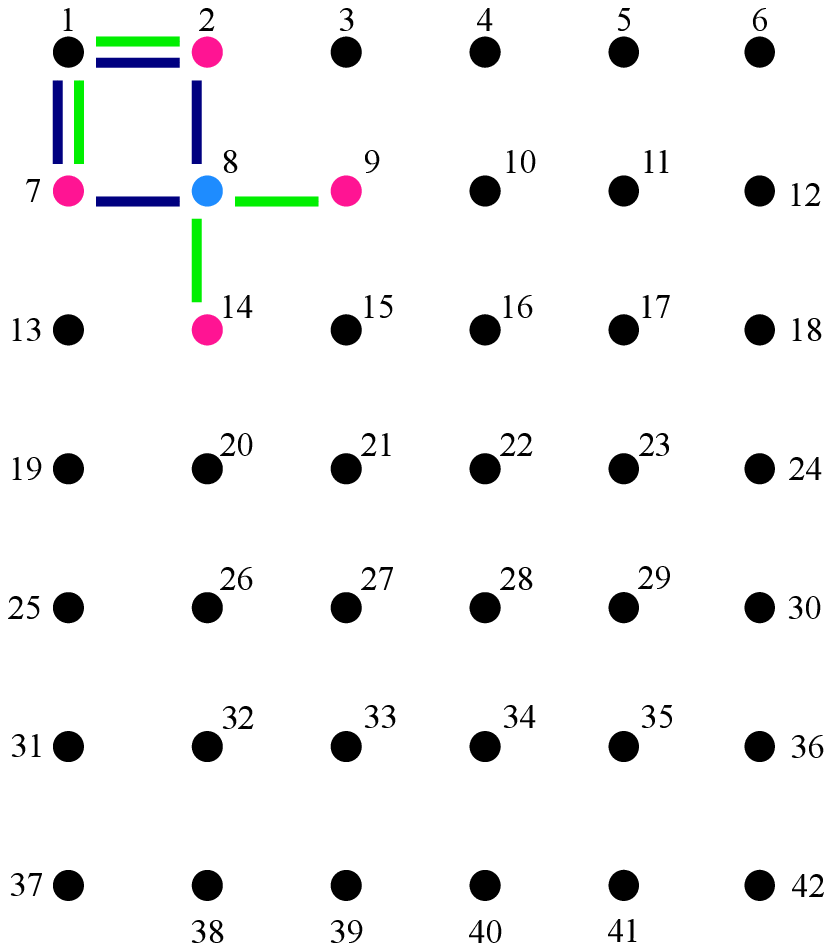}
\label{six2}}
\subfigure[]
{\includegraphics[width=4.8cm]{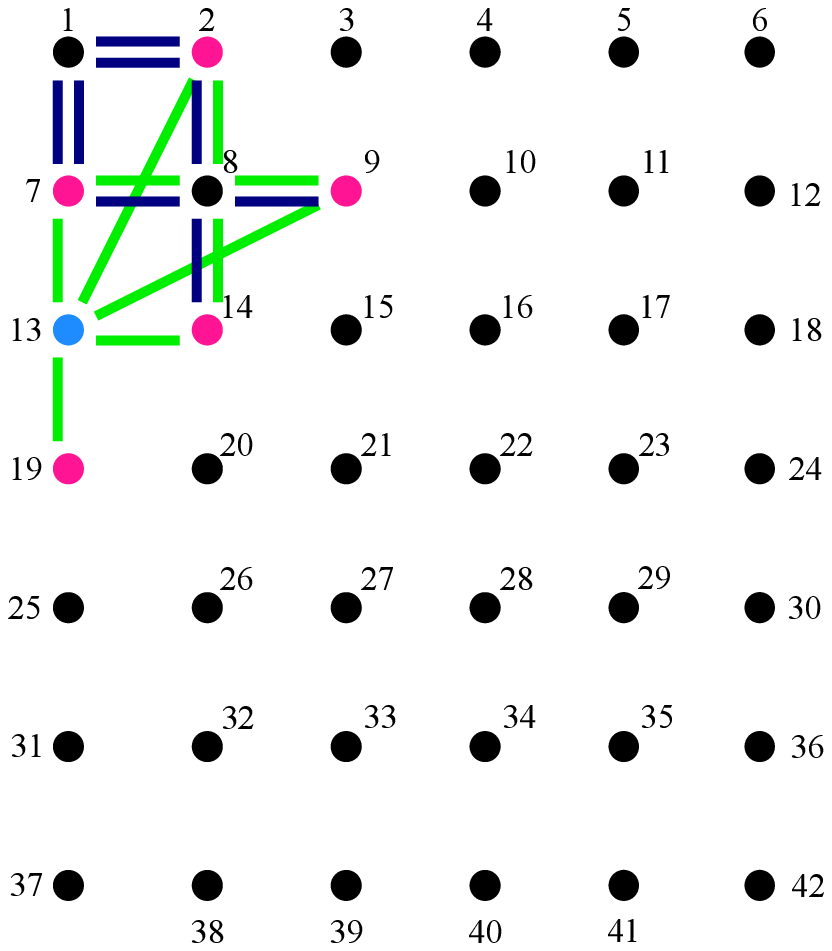}
\label{six3}}
\subfigure[]
{\includegraphics[width=4.8cm]{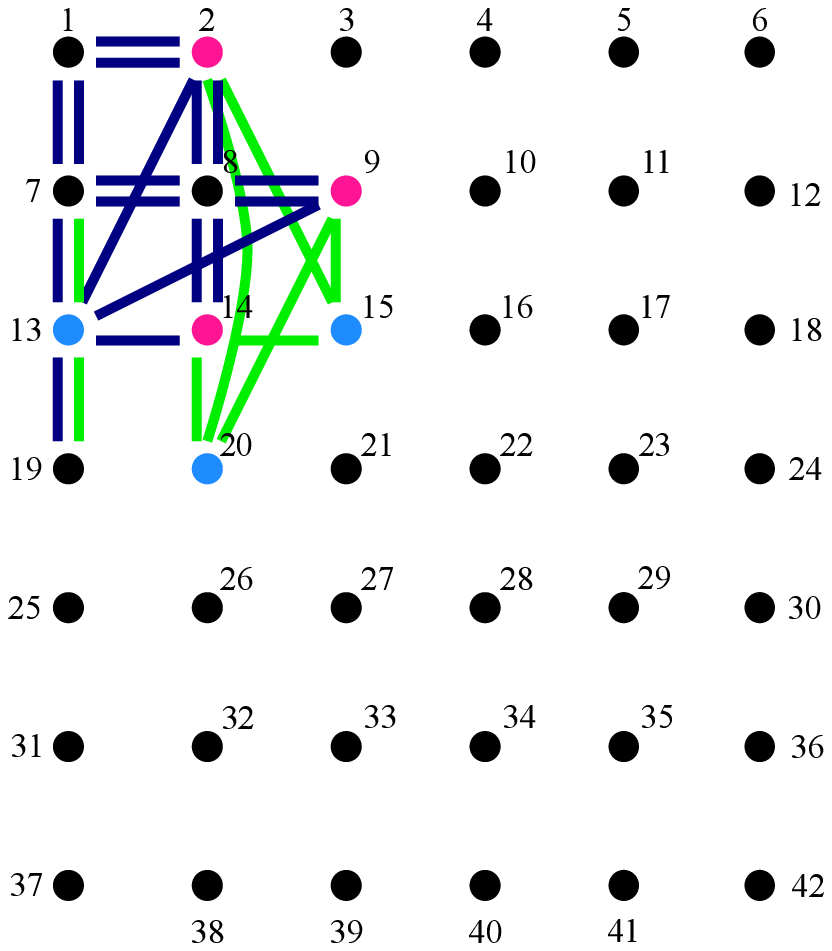}
\label{six4}}
\subfigure[]
{\includegraphics[width=4.8cm]{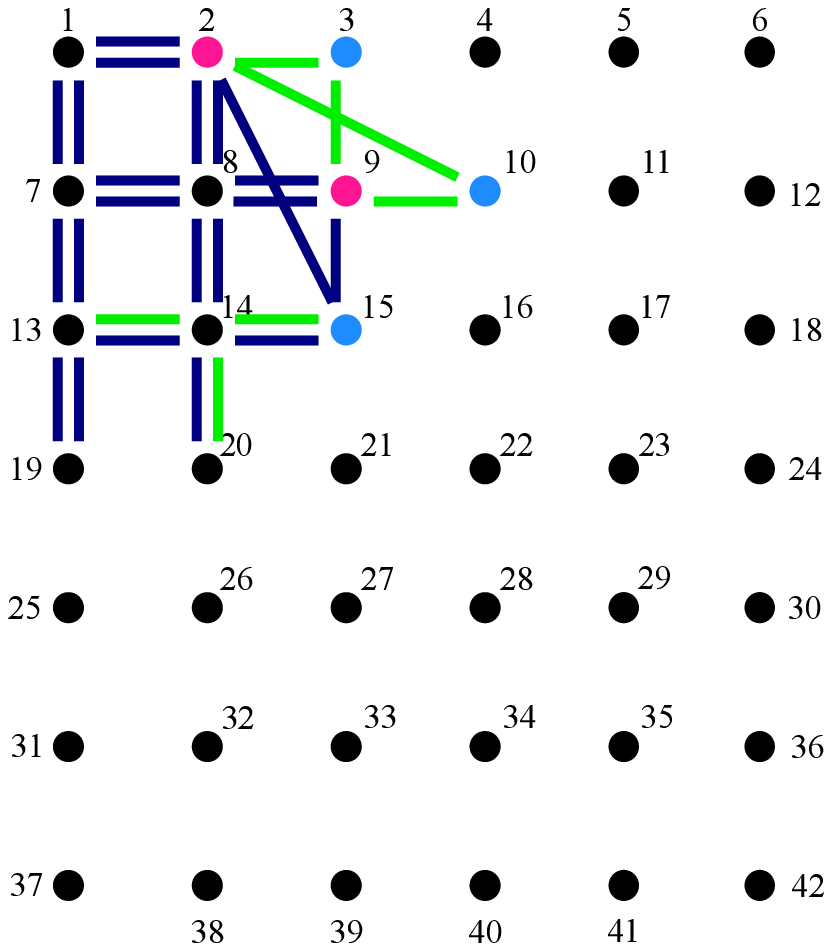}
\label{six5}}
\subfigure[]
{\includegraphics[width=4.8cm]{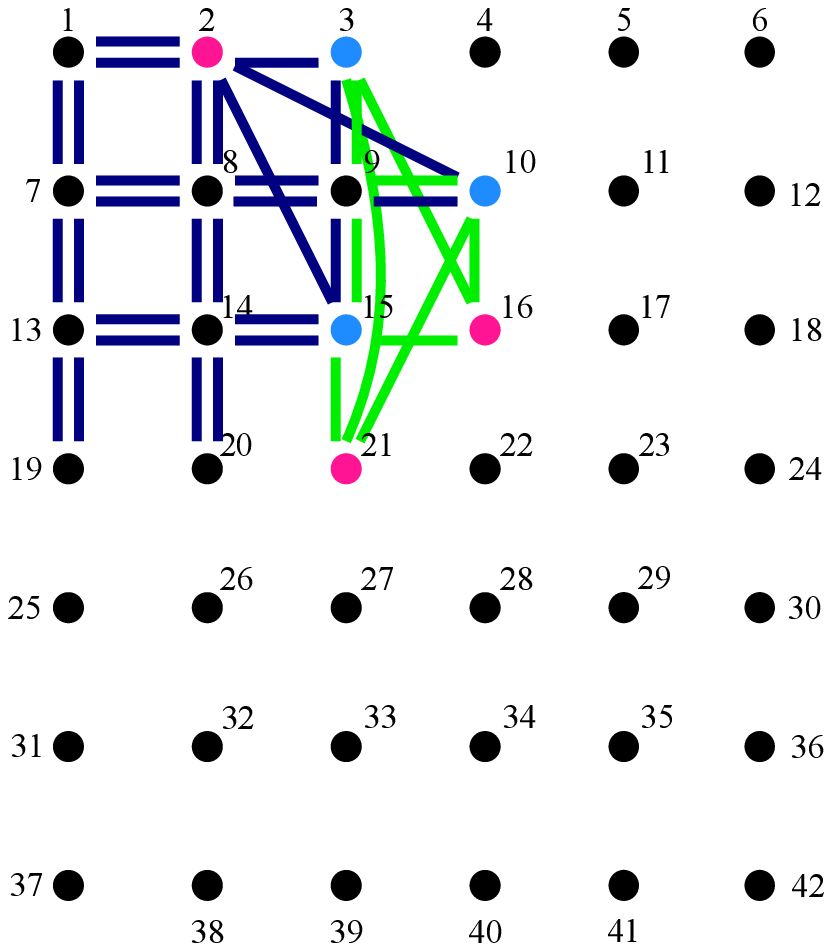}
\label{six6}}
\subfigure[]
{\includegraphics[width=4.8cm]{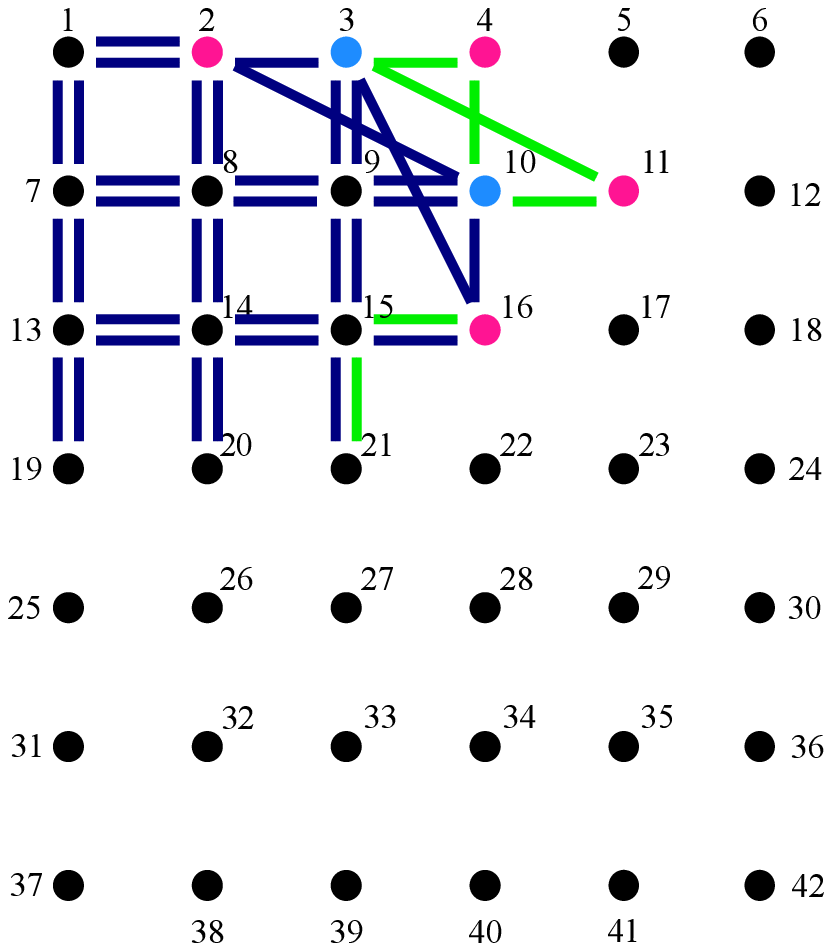}
\label{six7}}
\subfigure[]
{\includegraphics[width=4.8cm]{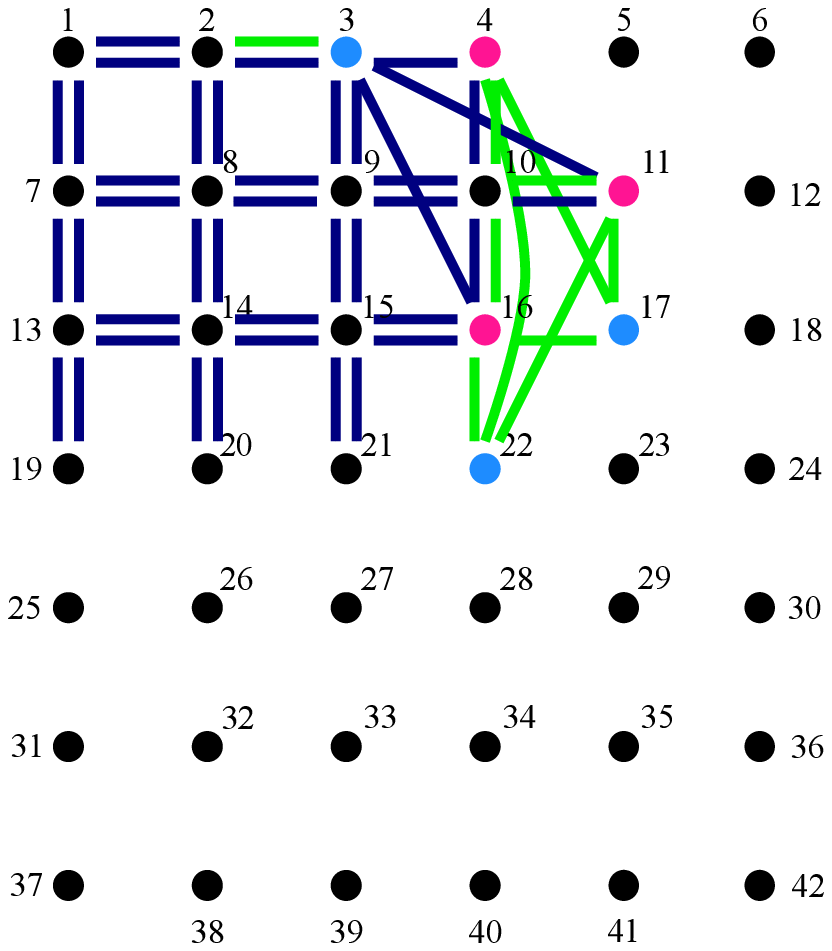}
\label{six8}}
\subfigure[]
{\includegraphics[width=4.8cm]{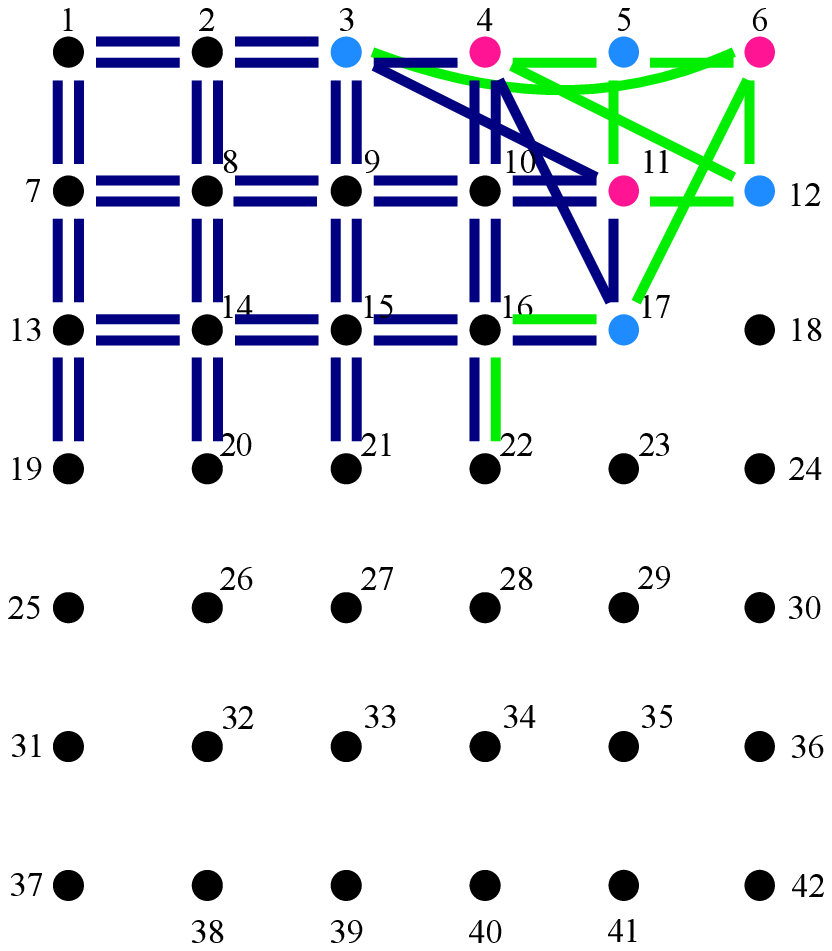}
\label{six9}}
\subfigure[]
{\includegraphics[width=4.8cm]{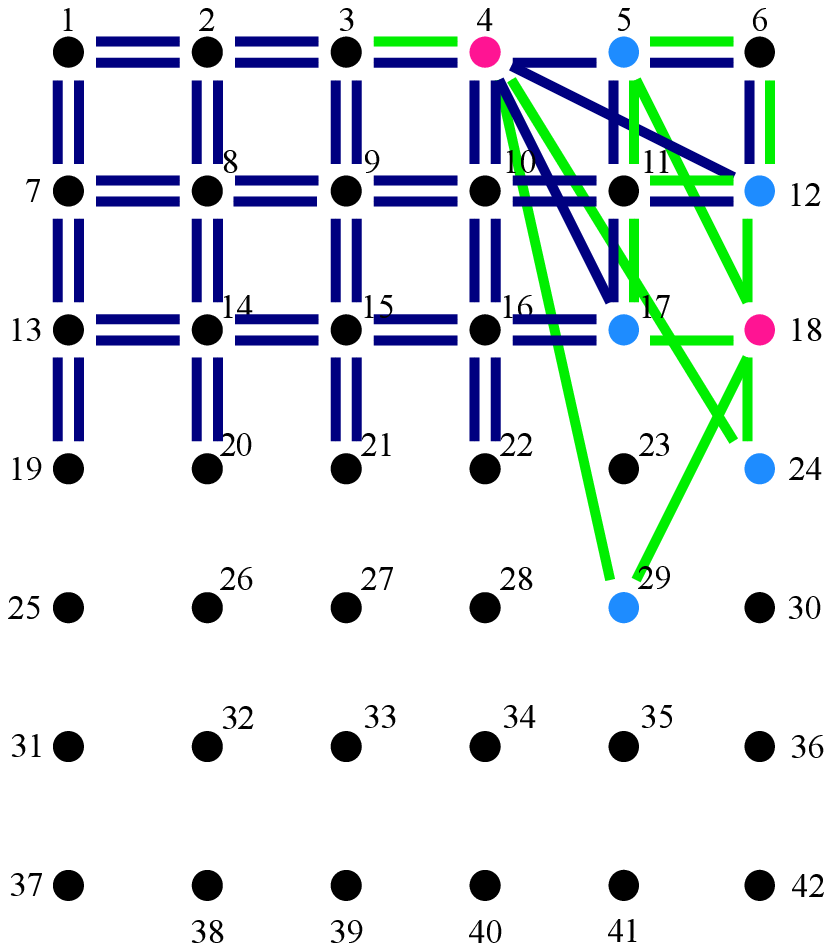}
\label{six10}}
\subfigure[]
{\includegraphics[width=4.8cm]{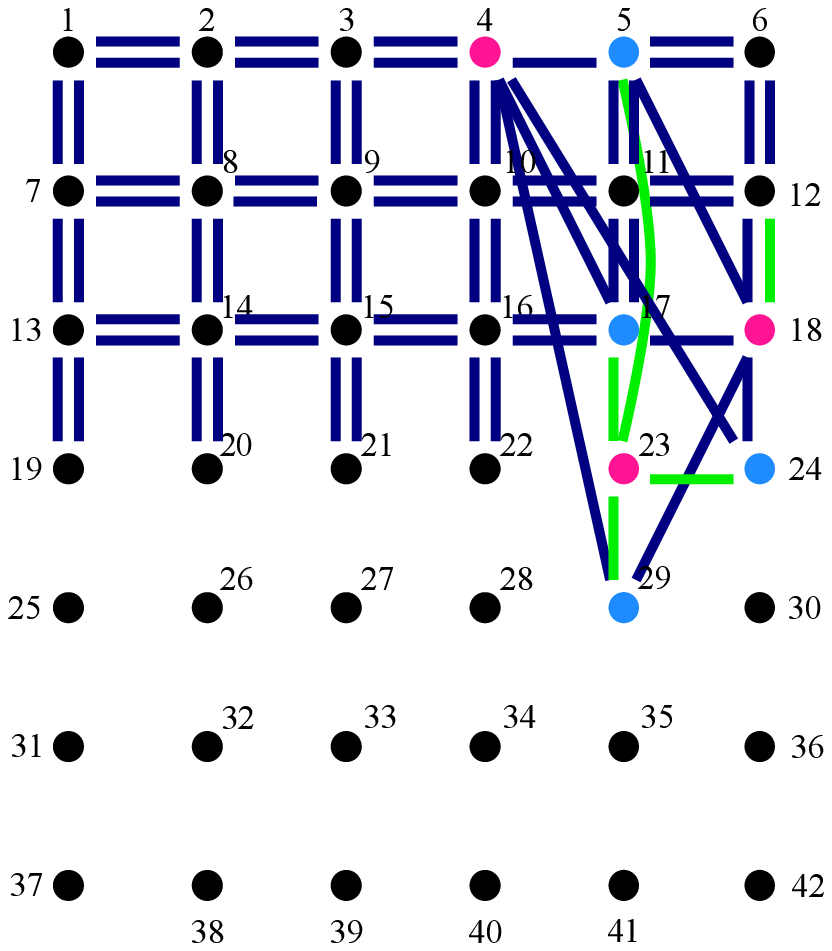}
\label{six11}}
\subfigure[]
{\includegraphics[width=4.8cm]{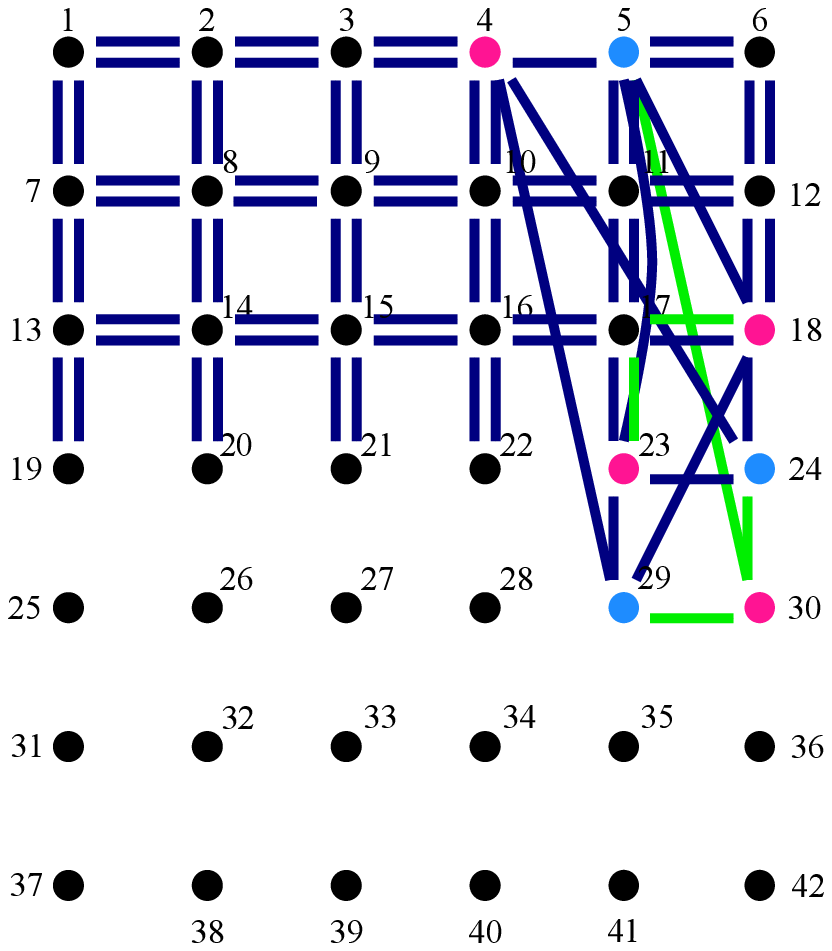}
\label{six12}}
\end{figure}

\end{document}